\documentclass[paper=a4, fontsize=11pt]{scrartcl} 
\usepackage[T1]{fontenc} 
\usepackage[english]{babel} 
\usepackage{amsfonts,amsthm} 
\usepackage{caption}
\usepackage{subcaption}
\usepackage{amsmath,mathtools}
\usepackage{graphicx}
\usepackage{float} 
\usepackage{blkarray} %
\usepackage{bm}
\usepackage{algorithm,algorithmic}
\usepackage{rotating}
\usepackage[super]{nth}
\usepackage[toc,page]{appendix} 
\usepackage{sectsty} 
\usepackage{tikz}
 \usetikzlibrary{patterns}
\usetikzlibrary{arrows} 
\usetikzlibrary{decorations.markings}
\usepackage{listings}
\usepackage{booktabs,caption,dcolumn,fixltx2e}
\newcolumntype{d}[1]{D{.}{.}{#1}}

\usepackage[utf8]{inputenc} 
\usepackage{blkarray}
\usepackage[english]{babel}
\newtheorem{theorem}{Theorem}[section]

\newtheorem{lemma}[theorem]{Lemma}
\newtheorem{exmp}{Example}[section]

\newtheorem{defn}{Definition}[section]

\usepackage[flushleft]{threeparttable}
\usepackage{enumitem}

\allsectionsfont{\centering \normalfont\scshape} 
\usepackage{array} 
\usepackage{fancyhdr} 
\newsavebox{\smallblockbox}

\makeatletter
\newcommand{\ALOOP}[1]{\ALC@it\algorithmicloop\ #1%
  \begin{ALC@loop}}
\newcommand{\ENDALOOP}{\end{ALC@loop}\ALC@it\algorithmicendloop}

\makeatother

\newcounter{algok}
\newtheorem{algo}[algok]{Algorithm}
\numberwithin{equation}{section} 
\numberwithin{figure}{section} 
\numberwithin{table}{section} 

\title{	
\normalfont \normalsize 
\huge Topological Techniques in Model Selection \\ 
}

\author{Shaoxiong Hu, Hugo Maruri-Aguilar, Zixiang Ma} 

\date{\normalsize\today} 

\begin{document}\maketitle 

\begin{abstract}
The LASSO is an attractive regularisation method for linear regression that combines variable selection with an efficient computation procedure. This paper is concerned with enhancing the performance of LASSO for square-free hierarchical polynomial models when combining validation error with a measure of model complexity. The measure of the complexity is the sum of Betti numbers of the model which is seen as a simplicial complex, and we describe the model in terms of components and cycles, borrowing from recent developments in computational topology. We study and propose an algorithm which combines statistical and topological criteria. This compound criteria would allow us to deal with model selection problems in polynomial regression models containing higher-order interactions. Simulation results demonstrate that the compound criteria produce sparser models with lower prediction errors than the estimators of several other statistical methods for higher order interaction models.
\end{abstract}

\section{Introduction}

\subsection{Modelling higher-order interactions}
Suppose we have $Y=(y_1,\cdots,y_n)^T\in \mathbb{R}^n$ is a set of $n$-dimensional observation vector, $X=(\mathbf{x}_1,\cdots,\mathbf{x}_p)\in \mathbb{R}^{n\times p}$ is the design matrix, $\theta \in \mathbb{R}^p$ is the unknown but fixed vector of parameters we wish to estimate and $\epsilon \in \mathbb{R}^n$ the vector of all independent error terms with zero mean and constant variance $\sigma^2$.
LASSO proposed by Tibshirani (1996) \cite{T1996} is an innovative variable selection method for regression. In particular, the LASSO uses a tuning parameter $\lambda$ as a weight that regulates how strongly $||\theta||_1$, which is the sum of the absolute values of the elements of $\theta$, penalises the least squares criterion. That is the penalised criterion function
\begin{equation}\label{equation2}
\frac{1}{2n}||Y-X\theta||_2^2+\lambda||\theta||_1.
\end{equation}
where $\parallel \cdot \parallel_2^2$ represents $L^2$-norm, $\parallel  \cdot \parallel_1$ is the sum of absolute values and $\lambda \in [0, \infty)$. LASSO continuously shrinks coefficients towards $0$ as $\lambda$ increases, in particular some coefficients are shrunk to $0$ when $\lambda$ is large enough (e.g. for $\lambda > 2\max_i \frac{1}{n}|\mathbf{x}^T_iY|$). 

The LASSO has recently attracted attention in the context of models with hierarchy restrictions. In these models, an interaction term is allowed only if both main effects are active (strong hierarchy) or if at least one main effect is active (weak hierarchy) \cite{Y1978, HW1992, C1996, J2006}, defined as follows:

Strong Heredity: If an interaction term is included in the model, then both of the corresponding main effects must be present.

Weak Heredity: If an interaction term is included in the model, then at least one of the corresponding main effects must be present.

For example, under strong hierarchy appearance of the term $\mathbf{x}_1\mathbf{x}_2$ in a model requires both $\mathbf{x}_1$ and $\mathbf{x}_2$, while under weak hierarchy at least one of $\mathbf{x}_1,\mathbf{x}_2$ is needed. The discussion for hierarchical polynomial models existed a long time ago, for example McCullagh (1984) \cite{M1984} said such constraints facilitate model interpretation and this can improve statistical power Cox (1984) \cite{CD1984}. Peixoto (1987) in \cite{PL1987} refers to hierarchical models as "well-formulated polynomial regression models". 

There is a number of paper considering fitting interaction models under strong or weak heredity constraints. For example, references including \cite{PL1987, F1991, PH2008, WD2010} apply constraints to enforce heredity to traditional step-wise model selection techniques \cite{C1996} and Bayesian methods proposed in \cite{GM1993}. Important recent contributions are the convex relaxation of Bien et al (2013) \cite{BTT2013} and Haris (2016) \cite{HW2016}. It considers a regression model for an outcome variable $Y$ and predictors $\mathbf{x}_1,\cdots,\mathbf{x}_p$, with pairwise interactions between these predictors. In particular, the model has the form linear regression where $X=(1, \mathbf{x}_1,\cdots,\mathbf{x}_p, \mathbf{x}_1\mathbf{x}_2,\cdots,\mathbf{x}_{p-1}\mathbf{x}_p)$ as in (\ref{pair_term}).
\begin{align}\label{pair_term}
 Y = \sum_j \theta_j \mathbf{x}_j + \frac{1}{2} \sum_{j \neq k} \theta_{jk}\mathbf{x}_j \mathbf{x}_k + \epsilon
\end{align}
The additive part $(\mathbf{x}_1,\cdots, \mathbf{x}_p)$ is referred as the 'main effect' terms and the quadratic part as the 'interaction' terms. They implemented strong and weak hierarchy versions of LASSO in their package \texttt{hierNet} and \texttt{FAMILY} which include the quadratic terms $\mathbf{x}^2$ as well. However, both strong and weak hierarchy only consider the case of pairwise interactions. In practice, there may potentially exist active interactions among three or even more variables. 

The polynomial regression model in our context can be formulated as follows 
\begin{equation}\label{polire}
Y = X\theta_{\alpha} + \epsilon
\end{equation} 
where $X=(\mathbf{x}^{\alpha})$, $\mathbf{x}^{\alpha}=\mathbf{x}_1^{\alpha_1}\cdots\mathbf{x}_p^{\alpha_p}$ is the regression term, $\alpha = (\alpha_1,\ldots,\alpha_p)$ and $\alpha_i \in \{0,1\}, i=1,\ldots,p$ is the exponent indicating the degree of the interactions. For example, $\mathbf{x}^{\alpha}$ becomes the linear term $\mathbf{x}_i$ when $\alpha_i = 1$ and $\alpha_j = 0$ for $j \neq i$. Moreover, $\mathbf{x}^{\alpha} = \mathbf{x}_i\mathbf{x}_j$ is the pairwise interaction term between $\mathbf{x}_i$ and $\mathbf{x}_j$ if $\alpha_i = \alpha_j = 1$ and $0$ for others. Here we did not put intercept term because it is always centred as zero in our context. In our work, we focus on square-free interactions. In other words, the exponent $\alpha_i$ is either $1$ or $0$. For example, there is no such terms $\mathbf{x}_i^2$, $i=1,\cdots,p$ in our regression model. We construct the higher interaction terms from the main effects and apply LASSO to the augmented data matrix $X$ which containing the main effects and the interactions. Coefficients are plotted versus the shrinkage factor $\lambda$. Efron et al. (2004) \cite{ET2004} suggested a modified LARS algorithm to determine the exact piecewise linear coefficients paths for the LASSO. They proved that the coefficients are piece-wise linearly along the path. In our context, we refer the piece-wise linear realization of coefficients as LASSO path and the points of intersection between two pieces as break points

Due to the complexity of the model terms in our situation, we consider to represent the hierarchical regression model in terms of topological objects. Graphic models was applied to represent statistical models in Meinshausen and B\"uhlmann (2006) \cite{MB2006} and Bien et al (2013) \cite{BTT2013} use vertices and edges to represent variables and pairwise interactions respectively for showing the sparsity pattern of the strong hierarchical LASSO. We extend the ideal to further step in higher-order interaction situations and consider to represent the regression model terms in terms of simplicial complex which is a set composed of vertices, edges, triangles and corresponding higher dimensional counterparts and closed for the subset belongs to the simplicial complex. Therefore a simplicial complex has a similar hierarchical structure as the hierarchical model we mentioned above.

This model representation provides us a way to link hierarchical regression models to homology. The progress in the area of topological modelling in statistics has extended the range of theoretical and applied problems studied with algebraic techniques. Persistent homology (PH) is a method used in topological data analysis (TDA) to study qualitative features of data that persist across multiple scales,  see Otter et al (2017) \cite{O2017} for a recent overview with an emphasis on publicly available software. 

\subsection{Organization of article}
In this paper, we show that a three-stage procedure could be used to deal with high order interaction problems. This method tends to have better prediction errors and recover better sparsity pattern compared to traditional statistical methods in both simulations and real data applications. This paper is organized as follows: Section 2 introduce the relevant background about homology groups which would be used to compute the number of independent cycles in the simplicial complex. The link between statistical models and Betti numbers developed in Section 2 has been exploited in Section 3 and we consider to represent regression model terms by simplicial complex. In Section 4, we illustrate that the change process of statistical models is consistent with the corresponding Betti numbers. In Section 5, we propose an algorithm combining statistical and algebraic criteria for model selection. We introduce the model errors and two other two statistical methods in Section 6 for the comparison of simulations.
We conduct several type of simulations to illustrate the performance of our algorithm in Section 7. An application to red wine quality data is in Section 8. The conclusion is in Section 9.

\section{Homology}

In this section, we presents elementary background for homology which provide a good foundation for handling Betti numbers. Homology is a mathematical formalism for telling us how a space is connected in a quantitative and unambiguous manner \cite{EH2010}. Homology groups provide a mathematical language for the holes in a topological  space. Instead of capturing holes directly, homology groups focus on what surrounds them. The cores of homology groups are group operations and maps that link topologically meaningful subsets of a space to each other. Here we describe some results due to \cite{EH2010} who introduce homology groups for computational topology which provides us a way to compute Betti numbers.

\subsection{Homology groups}
In our context, we use simplicial complexes as the prime objects to represent topological spaces. We briefly recall the definition of $d$-$simplex$. Suppose $x_1,x_2,\ldots,x_{d+1}\in \mathbb{R}^m$ are affinely independent points which means 
$\{x_{2}-x_{1}, x_{3}-x_{1},\dots ,x_{d+1}-x_{1}\}$ are linearly independent. We say 
$x=\sum_{i=1}^{d+1}\lambda_ix_i$ is a convex combination with $\lambda_i \in \mathbb{R}_{\geq 0}$.
The $k$-simplex determined by $x_1,x_2,\ldots,x_{d+1}$ is the set of points 
\begin{align*}
\{\lambda_1 x_1+\cdots+\lambda_{d+1}x_{d+1} \mid \sum_{i=1}^{d+1} \lambda_i=1\  \mbox{and}\ \lambda_i\geq 0\ \mbox{for  all of } i       \}
\end{align*}

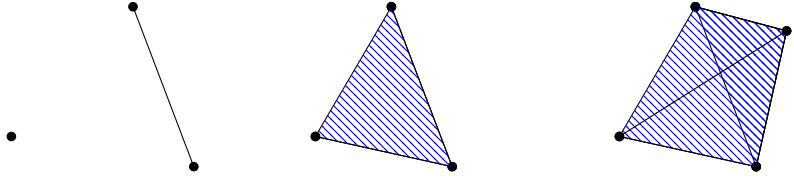
\begin{figure}
\centering%
\begin{tikzpicture}[scale=2,every label/.style={color=blue}]
\coordinate (1) at (0,0);
\draw [fill] (1)circle[radius=.8pt];
\coordinate  (11) at (0.8,0.86);
\coordinate  (13) at (1.2,-0.2);
\draw [fill] (11)circle[radius=.8pt] -- (13) circle[radius=.8pt];

\coordinate  (21) at (2.5,0.86);
\coordinate  (24) at (2,0);
\coordinate  (23) at (2.9,-0.2);
\draw [fill] (21)circle[radius=.8pt] -- (24) circle[radius=.8pt];
\draw [fill] (24)circle[radius=.8pt] -- (23) circle[radius=.8pt]; 
\draw [fill] (21)circle[radius=.8pt] -- (23) circle[radius=.8pt];
\draw[pattern=north west lines, pattern color=blue](21) -- (23) -- (24);

\coordinate  (x1) at (4.5,0.86);
\coordinate  (x2) at (5.1,0.7);
\coordinate  (x3) at (4.9,-0.2);
\coordinate  (x4) at (4,0);
\draw [fill] (x1)circle[radius=.8pt];
\draw [fill] (x1)circle[radius=.8pt] -- (x2) circle[radius=.8pt];
\draw [fill] (x2)circle[radius=.8pt] -- (x3) circle[radius=.8pt]; 
\draw [fill] (x3)circle[radius=.8pt] -- (x4) circle[radius=.8pt];
\draw [fill] (x1) -- (x4) circle[radius=.8pt];
\draw [fill] (x1)circle[radius=.8pt] -- (x3) circle[radius=.8pt];
\draw[dotted] [fill] (x2)circle[radius=.8pt] -- (x4) circle[radius=.8pt];
\draw[pattern=north west lines, pattern color=blue](x2) -- (x3) -- (x4);
\draw[pattern=north west lines, pattern color=blue](x1) -- (x2) -- (x3);
\draw[pattern=north west lines, pattern color=blue](x1) -- (x2) -- (x4);
\end{tikzpicture}
\caption{From left to right: a vertex, an edge, a triangle, and a tetrahedron. An edge has two vertices, triangle has three edges, and a tetrahedron has four triangles as faces.}
\label{figure1}
\end{figure}

We use the notation 
$\sigma= [x_1,x_2,\ldots,x_{d+1}]$ 
to represent $d$-simplex. Its dimension is $\mbox{dim}(\sigma)=d$. We use special names for the fist few dimensions: $vertices$ for 0-simplex, $edge$ for 1-simplex, $triangle$ for 2-simplex, and $tetrahedron$ for 3-simplex
as in Figure \ref{figure1}.
Now we start to transform our intuitions of chains and cycles (holes) into topological language. Let $\Delta$ be a simplicial complex which is a set of simplices that satisfies the two conditions: Every subset of a simplex from $\Delta$ is also in $\Delta$ and the intersection of any two simplices $\sigma_1 \in \Delta, \sigma_2 \in \Delta$ is a subset of both $\sigma_1$ and $\sigma_2$. Cycles are the special case of the \textit{$d$-chain} which defined as follows.
\begin{defn}\label{chain}
A $d$-chain is a formal sum of $d$-simplices in $\Delta$. The notation for this is $c=\sum a_i\sigma_i$, where $\sigma \subset \Delta$ is a $d$-simplex and the coefficients $a_i$ are either $0$ or $1$, called $modulo\ 2$ coefficients. 
\end{defn}
In the above definition, we use formal sum to represent the existence of a certain $d$-simplex. Geometrically, we might think of the union of two points $x_1$ and $x_2$ as being the sum of two $0$-simplex $[x_1]+[x_2]$.  From the Definition \ref{chain}, a $d$-chain consists of all $d$-simplices in the simplicial complex $\Delta$ and the coefficients $a_i$ means the attendance of the simplex $\sigma_i$, in specific, if $a_i=1$, this means $\sigma_i$ is in $\Delta$, otherwise there is no such $p$-simplex in $\Delta$. We also find that two d-chains are added component-wise. Specially, suppose we have $c=\sum a_i\sigma_i$ and $c'=\sum b_i\sigma_i$, then $c+c'=\sum (a_i+b_i)\sigma_i$, where the coefficients satisfy $1+1=0$. This addition operation is very important in describing the idea of different dimensional cycles (holes) in the simplicial complex. Since the set of all $d$-chain is closed under the addition operation, the $d$-chains together with the addition operation form the $group \ of\ d-chains$ denoted as $C_d=C_d(K)$. Moreover, the identity element of $C_d$ is $0=\sum 0\sigma_i$. It is not difficult to see that $c+c=0$, so the inverse of $c$ is $-c=c$. Finally, $C_d$ is commutative (abelian) because addition modulo 2 is commutative. We point out that $C_d$ is zero when $d$ is less than zero or greater than the dimension of $\Delta$. 

Since the homology groups describe the cycles by focusing on the type of boundaries that border the cycles, we define the boundary of a $d$-simplex as the sum of its $(d-1)$-dimensional faces. 

\begin{defn}\label{defn1}
Let $\sigma=[x_1,\ldots,x_{d+1}]$ be the $d$-simplex spanned by the listed vertices $x_1,\ldots,x_{d+1}$, its boundary is 
\begin{align} \label{bound}
\partial_d \sigma = \sum_{j=1}^{d+1}[x_1,\ldots,x_{j-1}, x_j, \ldots,x_{d+1}]
\end{align}
where $[x_1,\ldots,x_{j-1}, x_j, \ldots,x_{d+1}]$ is the $(d-1)$-simplex spanned by vertices $x_1,\ldots,x_{j-1},$ $x_j, \ldots,$ $x_{d+1}$.
\end{defn}
For illustration, the boundary of a triangle $[x_1,x_2,x_3]$ depicted in Figure \ref{figure1} is $[x_2,x_3]+[x_1,x_3]+ [x_1,x_2]$.
For a $d$-chain, $c=\sum a_i\sigma_i$, by the Definition \ref{defn1}, the boundary is the formal sum of the boundaries of its simplices, 
$\partial_d c= \sum a_i\partial_d\sigma_i$. 
This means $\partial_d c$ is a $(d-1)$-chain. In other words, taking the boundary maps a $d$-chain to a $(d-1)$-chain, and this can be represented as: 
$\partial_d: C_d \rightarrow C_{d-1}$. We also notice that taking the boundary commutes with addition operations, to be specific, $\partial_d(c+c')=\partial_d c+ \partial_d c'$. This is again the defining property of a \textit{homomorphism}  which is a map between groups that commutes with the group operation. Therefore we can construct the chain complex which is the sequence of chain groups connected by boundary homomorphisms,
\[
\cdots
\xrightarrow{\partial_{d+2}}
C_{d+1}
\xrightarrow{\partial_{d+1}}
C_d
\xrightarrow{\partial_d }
C_{d-1}
\xrightarrow{\partial_{d-1}}
\cdots
\]

With the $d$-chain introduced above, we now focus on two particular types of chains and use them to characterize the cycles (holes) and the corresponding boundaries. Intuitivly, there is no isolated end point in a cycle (say an unfilled triangle) since each end point is connected to from the hole. Therefore a good way to define a cycle is to described it as a chain but with empty boundary since the formal sum are modulo 2 coefficients \cite{EH2010}.

\begin{defn}\label{defn2}
A $d$-cycle is a $d$-chain with empty boundary, $\partial c=0, c \in C_d$. The set of all $d$-cycles together with addition operation forms a group since $\partial$ commutes with addition, denoted as $Z_d:=Z_d(\Delta)$.
\end{defn}
From Definition \ref{defn2}, the $d$-cycle group $Z_d$ is a subgroup of $C_d$ because each element in $Z_{d}$ has zero boundary. Moreover, the group of $d$-cycle is the kernel of the $d$\textsuperscript{th} boundary homomorphism, $Z_d=\mbox{ker} (\partial_d)$.

An another important object for distinguishing different dimensional cycles is the boundary of the $d$-chain.
\begin{defn}
A $d$-boundary is a $d$-chain that is the boundary of a $(d+1)$-chain, $c=\partial_d$ with $d\in C_{d+1}$. 
In the same way as $d$-cycles, we have a 
group of $d$-boundaries, denoted as $B_d=B_d(\Delta)$, which is again a subgroup of the $d$-chains. 
\end{defn}
From the definition of $d$-boundary, the group of $d$-boundaries is the image of the $(d+1)$\textsuperscript{th} boundary homomorphism, $B_d=\mbox{im}(\partial_{d+1})$. Since the chain groups are abelian, so are their boundary subgroups.

The following lemma \cite{EH2010} indicates that the boundary of boundary is necessarily zero which makes homology work and provide solid support for the classification of cycles of simplicial complexes.

\begin{lemma}\label{lemma2}
\textbf{Fundamental Lemma of Homology \cite{EH2010}.}\ The boundary of the boundary is zero, i.e. $\partial_d\partial_{d+1}f=0$ for every integer $d$ and every $(d+1)$-chain $f$.
\end{lemma}

It follows from Lemma \ref{lemma2} that every $d$-boundary is also a $d$-cycle since every $d$-boundary must be the boundary of a $d+1$-chain, in other words, $B_d$ is a subgroup of $Z_d$. 

We can take quotients since the boundaries form subgroups of the cycle groups. To be specific, we can partition each cycle group into classes of cycles that differ from each other by boundaries. This leads to the concepts of homology groups and their ranks which we would introduce in this section. This also gives us an alternative view of Betti numbers which is exactly the rank of the corresponding homology group being used to describe the number of independent cycles of the simplicial complex.

In group theory, the rank of a group refers to the minimal number of generators of this group defined as follows.
\begin{defn}\label{defn4}
The rank of a group $G$, denoted rank($G$) is the smallest cardinality of a generating set for $G$, that is 
\begin{align*}
\mbox{rank}(G)=\mbox{min}\ \{\mid X\mid: X\subset G, \langle X\rangle=G\}
\end{align*}
where $X$ is minimal subset of $G$ that could generate $G$.
\end{defn}

We introduce the definition of homology group which consists of independent cycles with the same type of boundaries.
\begin{defn}
\label{defn3}
The $d$\textsuperscript{th} homology group is the $d$\textsuperscript{th} cycle group modulo the $d$\textsuperscript{th} boundary group, $H_d(\Delta)=Z_d/ B_d$. Here we write $H_d=H_d(\Delta)$ for short. The $d$\textsuperscript{th} Betti number is the rank of this group, $\beta_d=\mbox{rank}(H_d)=\mbox{rank}(Z_d)- \mbox{rank}(B_d)$. 
\end{defn}

Every element of the $d$\textsuperscript{th} homology group $H_d$ is obtained by adding all $d$-boundaries to a given $d$-cycle, $c+B_d$ with $c\in Z_d$. If there is another cycle $c'=c+c''$, with $c''$ an element of $B_d$, we have that $c$ and $c'$ are in the same class, $c'+B_d=c+B_d$ because $c''+B_d=B_d$.

\begin{defn}
Any two cycles in the same homology class are said to be homologous.
\end{defn}

Similarly, if $c$ and $c'$ are from two classes, 
\begin{align*}
(c+B_d)+(c_0+B_d)=(c+c_0)+B_d
\end{align*}
is closed under the addition operation. We thus see that $H_d$ is indeed a group. Moreover, $H_d$ is a abelian since $Z_d$ is abelian.

The rank of the homology group is the rank difference between the corresponding cycle group and the boundary group which gives us the number of classes of cycles surrounded by the same dimensional boundaries. In other words, the Betti number of the homology group represent the number of independent cycles (holes) in the simplicial complex which is often less than the number of elements in the corresponding cycle group. Next we introduce a special type of simplicial complex which consists of same dimensional simplexes. 

\begin{defn}\label{d-complete}
A simplicial complex $\Delta$ with $m$ vertices is $d$-closed if the set of maximal faces of $\Delta$ contains all possible $d$-simplex, $d\geq 1$. Denote $d$-dimensional $d$-closed simplical complex with $m$ vertices as $\Omega^d_m, m\geq d+2$.
\end{defn}
For the simple cases, $\Omega^1_3$ is a triangle with three edges which is the boundary of $2$-simplex and $\Omega^2_4$ is a hollow tetrahedron which is the boundary of $3$-simplex. The following theory states that the smallest $d$-cycle is the boundary of a $(d+1)$-simplex.
\begin{theorem}\label{prop1}
A $d$-cycle can have no fewer than $d+2$ vertices. If $\sigma$ is a $d$-cycle on $d+2$ vertices, then $\sigma = \Omega^d_{d+2}$ where $d\geq1$.
\end{theorem}

For a simplicial complex with $d+2$ vertices, the highest dimensional cycles contained in this simplicial complex can be only $d$-cycles since the smallest $d$-dimensional cycle is the boundary of a $(d + 1)$-simplex. This indicates that the Betti numbers of the simplicial complex is not an infinite sequence and we have $\beta_k = 0$ if $k>d$. Informally speaking, $k$-cycle has to be bounded by the corresponding $k$-simplices and it can not be a $k$-cycle since there are not enough $k$-simplices for the cycle's boundary.

The following lemma verifies that the number of independent cycles of the simplicial complex consisting of all edges is less than the number we observe by eyes.
\begin{lemma}\label{lemma4}
Let $\Delta$ be a connected simplicial complex with $d+2$ vertices ($d\geq1$), comprised of all possible edges formed by the vertices, then the number of independent $1$-cycles is
$\binom{d+1}{2}$, that is the Betti number $\beta_1 = \mbox{rank}\ (H_1)=\binom{d+1}{2}$.
\end{lemma}

The above conclusion can be extended to the general case. We show that the number of independent cycles of the simplicial complex formed by homogeneous simplices follows the pattern of Pascal's triangle as in Table \ref{pa2} but without the first and the second column of the original version. 
\begin{table}
\centering
\caption{Pascal's triangle for Betti numbers with $d+2$ vertices}
\label{pa2}
\begin{tabular}{>{$}l<{$}|*{7}{c}}
\multicolumn{1}{l}{$d$} &&&&&&&\\\cline{1-1} 
1 &&1&&&&&\\
2 &&3&1&&&&\\
3 &&6&4&1&&&\\
4 &&10&10&5&1&&\\
5 &&15&20&15&6&1&\\
6 &&21&35&35&21&7&1\\\hline
\multicolumn{1}{l}{} &&1&2&3&4&5&6\\\cline{2-8}
\multicolumn{1}{l}{} &\multicolumn{7}{c}{$k$}
\end{tabular}
\end{table}

\begin{theorem}\label{pa1}
The number of independent $k$-cycles of $d$-closed simplicial complex $\Omega^k_{d+2}$ is $\binom{d+1}{k+1}$, i.e. $\mbox{rank}\ (B_k)=\binom{d+1}{k+1}$. In order to guarantee the existence of  $k$ dimensional cycle, we require that $d\geq k\geq1$.
\end{theorem}

Similar to Lemma \ref{lemma4}, Theorem \ref{pa1} indicates that the same dimensional cycles in the simplicial complex are not necessary to be independent in the general case. The independent cycles form the generators of the corresponding cycle group and the rest cycles can be represented by those independent generators. Since simplicial complex $\Omega^k_{d+2}$ is connected, we note that $\mbox{rank}\ (B_0) = 1$.

\subsection{Computation of Simplicial Homology: An Algorithmic View}\label{2.2}
In this section, we will introduce an algorithmic way to compute Betti numbers by representing the boundary homomorphisms in terms of boundary matrices which entries are either zero or one. This transforms the rank of the cycle group to the rank of the kernel matrix and the rank of the boundary group becomes the rank of the corresponding image matrix. Therefore the calculation of homology with integer coefficients of a simplicial complex reduces to the calculation of the Smith Normal Form of the boundary matrices which in general are sparse \cite{DHS2003}. We provide a review of an algorithm to compute Betti numbers in \cite{DHS2003} which uses elementary operations of a matrix in linear algebra for the calculation of Smith Normal Form of sparse matrices.

For a given simpliclal complex $\Delta$ and any two $d$-simplexes $\sigma_1, \sigma_2\in \Delta$, we have $\partial_d(\sigma_1 + \sigma_2) = \partial_d \sigma_1 + \partial_d \sigma_2$ by (\ref{bound}) which means $\partial_d$ is a linear operator on $C_d$ over $\mathbb{Z}_2$. Since we have $Z_d = \mbox{ker}(\partial_d)$ and $B_{d-1} = im (\partial_d)$, by the Rank-nullity theorem in linear algebra \cite{MC2000} we have 
\begin{align*}
C_d \simeq Z_d \oplus B_{d-1}
\end{align*}
which also indicates the following import result
\begin{align}\label{de}
m_d = z_d + b_{d-1}
\end{align} 
where $m_d$ is the number of $d$-simplex in $\Delta$, $z_d$ and $b_{d-1}$ are the rank of $Z_d$ and $B_{d-1}$, namely $z_d = \mbox{rank}(Z_d)$ and $b_{d-1} = \mbox{rank}(B_{d-1})$. 

If we can decompose $m_d$ into (\ref{de}), we would be able to compute the $d$\textsuperscript{th} as $\beta_d = z_d - b_{d}$. Let $C_d = \mbox{span}\{\sigma_1,\ldots,\sigma_{m_d}\}$ and $C_{d-1} = \mbox{span}\{\tau_1,\ldots,\tau_{m_{d-1}}\}$, 
where $\sigma_i \in \Delta$, $i = 1,\ldots,m_d$ are the $d$-simplexes of $\Delta$ and $\tau_j \in \Delta$, $j = 1,\ldots, m_{d-1}$ are the $(d-1)$-simplices of $\Delta$. By \ref{bound} we have 
\begin{align}\label{re}
\partial_d \sigma_i = \sum_{j=1}^{m_{d-1}}a^i_j \tau_j
\end{align} 
where $a^i_j = 1$ if and only if $(d-1)$-simplex $\tau_j$ belongs to $\partial_d$, otherwise $a^i_j = 0$.

For all $\sigma_i, i = 1,\ldots, m_d$, we write the relation (\ref{re}) into matrix as follows
\begin{align}\label{matrix}
A_{\partial_d} = 
\begin{blockarray}{ccccc}
& \sigma_1 & \sigma_2 & \cdots & \sigma_{m_d} \\
\begin{block}{c(cccc)}
 \tau_1 & a^1_1 & a^2_1 & \cdots & a^{m_d}_1 \\
 \tau_2 & a^1_2 & a^2_2 & \cdots & a^{m_d}_2  \\
  \vdots & \vdots & \vdots & \ddots & \vdots  \\
\tau_{m_{d-1}} & a^1_{m_{d-1}} & a^2_{m_{d-1}} & \cdots & a^{m_d}_{m_{d-1}}  \\
 \end{block}
\end{blockarray}
\end{align}
where each column of $A_{\partial_d}$ represent the boundary coefficients of $\partial_{d}\sigma_i$ and $a^i_j$ is the $(j,i)$ entry of the boundary matrix (\ref{matrix}). Therefore $m_d$ is the number of columns of the matrix $A_{\partial_d}$ which is the total rank of the column space. Also we have $B_{d-1} = \mbox{im}(\partial_d) =$ the column space of $A_{\partial_d}$ so that $\mbox{rank}(B_{d-1}) = \mbox{im}(\partial_d) = \mbox{rank}(A_{\partial_d})$ with knowledge of linear algebra, here $\mbox{rank}(A_{\partial_d})$ we mean the rank of the column space of $A_{\partial_d}$ instead of the rank of the matrix. The boundary matrix $A_{\partial_d}$ encodes all the possible relationship between $d$-simplexes and the corresponding boundaries.

Let $c = \sum_{i=1}^{m_d}c_i \sigma_i$ be a $d$-chain in $\Delta$, where $c_i = 1$ if $\sigma_i$ belongs to $c$, otherwise $c_i = 0$. So the boundary of $c$
\begin{align}\label{c}
\partial_d c = \partial_d \sum_{i=1}^{m_d}c_i \sigma_i =  \sum_{i=1}^{m_d}c_i \partial_d\sigma_i = \sum_{i=1}^{m_d}c_i \sum_{j=1}^{d_{d-1}}a^i_j \tau_j = \sum_{i=1}^{m_d} (\sum_{j=1}^{m_{d-1}}a^i_j c_i) \tau_j
\end{align}
If we regard the $(d-1)$-simplexes $\{\tau_1, \ldots, \tau_{m_{d-1}}\}$ as a basis as in linear algebra, we have the coordinate of $\partial_d c$ in (\ref{c}) written into the matrix form as follows
 \[
\begin{blockarray}{ccccc}
\begin{block}{(cccc)(c)}
 a^1_1 & a^2_1 & \cdots & a^{m_d}_1 & c_1\\
  a^1_2 & a^2_2 & \cdots & a^{m_d}_2 & c_2\\
  \vdots & \vdots & \ddots & \vdots & \vdots\\
 a^1_{m_{d-1}} & a^2_{m_{d-1}} & \cdots  & a^{m_d}_{m_{d-1}}&   c_{m_d} \\
 \end{block}
\end{blockarray}
 \]

From the boundary matrix in (\ref{matrix}) above, we may consider to compute the rank of $B_{d-1}$ through the well-known linear algebra technique, Gaussian elimination. Now we formally introduce the algorithmic method to decompose $m_d$ as (\ref{de}).
In linear algebra, exchanging two columns or adding one columns to another do no change the rank of a matrix, so do the similar operations for the rows. Since we can write all relations between $d$-simplexes and $(d-1)$-simplexes of the simplicial complex $\Delta$ into a matrix $A_{\partial_d}$, we introduce four elementary operations of columns and rows on the boundary matrix which do not change the rank of the matrix $A_{\partial_d}$. Let $\tilde{a}^{i}_j$ be the new $(j, i)$ entry of the boundary matrix after the elementary operations.
\begin{enumerate}[label=(\alph*)]
\item Exchanging two columns, let $\tilde{a}^{i_1}_j = {a}^{i_2}_j$ and $\tilde{a}^{i_2}_j = {a}^{i_1}_j$ for $j = 1,\ldots,m_{d-1}$.
\item Adding one column to another, $\tilde{a}^{i_1}_j = {a}^{i_1}_j + {a}^{i_2}_j$.
\item Exchanging two rows, let $\tilde{a}^{i}_{j_1} = {a}^{i}_{j_2}$ and $\tilde{a}^{i}_{j_2} = {a}^{i}_{j_1}$ for $i = 1,\ldots,m_{d}$.
\item Adding one row to another, $\tilde{a}^{i}_{j_1} = {a}^{i}_{j_1} + {a}^{i}_{j_2}$.
\end{enumerate}  
It is not difficult to see that the four elementary operations of columns and row will not change the rank of the boundary matrix. We can reduce the boundary matrix $A_{\partial_d}$ into the following diagonal form by using the above four elementary operations.
\begin{align}\label{dia}
N_{\partial_d} =
\begin{blockarray}{cccccc}
\begin{block}{(cccccc)}
 1 &  &  &  &  &\\
  & \ddots  &  & &   \\
  & & 1 & & & &      \\
  & & & 0 & &        \\
  &  &  & & \ddots &  \\
  &   &   &  & & 0 \\
 \end{block}
\end{blockarray}
\end{align}
If we could reduce the boundary matrix into this diagonal form (\ref{dia}), the rank of $A_{\partial_d}$ is equal to the number of $1$s on the diagonal of the matrix $N_{\partial_d}$.
We introduce the process in \cite{EH2010} which is similar to Gaussian elimination for solving system of linear equations. We can move a $1$ to the upper left corner in at most two exchange operations. With at most $m_{d}-1$  column and $m_{d-1}-1$ row additions, we can cancel out the rest $1$s in the first row and first column. We then recurse for the submatrix obtained by removing the first row and first column.  We first initialize the matrix to $A_{\partial_d}[j,i] = a^i_j$ and let $x$ be the row and column number of upper left element of the sub-matrix we consider and be initialize to $1$. We summary the algorithm as in the following Algorithm \ref{algo}.

\begin{algorithm}
\caption{Reduce the boundary matrix $A_{\partial}$ to diagonal form $N_{\partial}$ from $x$}
\label{algo}
\begin{algorithmic}[1]
   \REQUIRE The boundary matrix $A_{\partial}$.
    \ENSURE The reduced diagonal form $N_{\partial}$.
 \STATE \COMMENT{\textbf{Step 1}} Exchange columns and rows such that $a^1_1 = 1$.
 \IF{$\exists$ $k\geq x$, $l\geq x$ with $a_l^k = 1$}
    \STATE exchange columns $x$ and $k$; exchange rows $x$ and  $l$;
    \STATE \COMMENT{\textbf{Step 2}} Eliminate other 1s in row $x$.
     \FOR{$j = x+1$ to $m_{d}$}
        \IF{$a^j_x = 1$}
           \STATE add column $x$ to column $j$
        \ENDIF 
     \ENDFOR ;
         \STATE \COMMENT{\textbf{Step 3}} Eliminate other 1s in column $x$.
     \FOR{$i=x+1$ to $m_{d-1}$}
       \IF{$a^x_i = 1$}
         \STATE add row $x$ to row $i$
       \ENDIF
     \ENDFOR ;
    \STATE REDUCE($x+1$)
 \ENDIF 
\end{algorithmic}
\end{algorithm}

Now we carefully go through this algorithm for $x = 1$ which means $a^1_1 = 1$. If there is no $1$s in other places, i.e., $a_l^k = 0$, then we are done. If there exists $a_l^k = 1$, we could move this $1$ to the upper left corner by exchanging the fist row and row $l$ and exchanging the fist column and column $k$ such that $\tilde{a}^1_1 = 1$. The next step is to eliminate the rest $1$s in the first row and the first column of the new matrix. This can be achieved by adding the first row to the rest rows thus cancelling out $1$s in the first column. And then adding the first column to the rest columns zeros out $1$s in the first row. We repeat this process recursively to the submatrix without the first row and first column. The final matrix left would be $N_{\partial_d}$.
We have at most $m_{d-1}$ row and $m_d$ column operations per recursive call and total $O(m_{d-1}m_{d}(m_{d-1}+m_{d}))$ complexity. For more details of the computational complexity of Betti numbers, see \cite{DE1995, EP2014}

\section{Models, Simplicial Complexes, Betti numbers}
In this section, we will explore the relationship between regression model terms and the Betti numbers. We would interpret the Betti number as the measurement of model complexity. In other words, the statistical model with smaller Betti numbers is of less complexity than the one with greater Betti numbers when both models have the same size. 

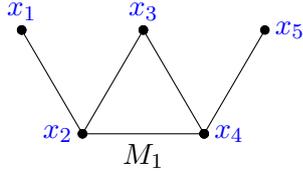
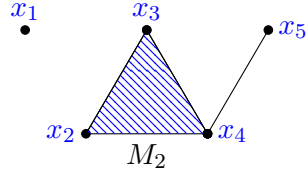
\begin{figure}
\centering
\begin{subfigure}[t]{0.45\textwidth}
\centering
\begin{tikzpicture}[scale=2,every label/.style={color=blue}]
\coordinate [label=above:$x_1$] (x1) at (-0.4,0.688);
\coordinate [label=left:$x_2$] (x2) at (0,0);
\coordinate [label=above:$x_3$] (x3) at (0.4,0.688);
\coordinate [label=right:$x_4$] (x4) at (0.8,0);
\coordinate [label=right:$x_5$] (x5) at (1.2,0.688);
\draw [fill] (x1)circle[radius=.8pt];
\draw [fill] (x1)circle[radius=.8pt] -- (x2) circle[radius=.8pt];
\draw [fill] (x2)circle[radius=.8pt] -- (x3) circle[radius=.8pt]; 
\draw [fill] (x3)circle[radius=.8pt] -- (x4) circle[radius=.8pt];
\draw [fill] (x2) -- node[below] {$M_1$}(x4) circle[radius=.8pt];
\draw [fill] (x4)circle[radius=.8pt] -- (x5) circle[radius=.8pt];
\end{tikzpicture}
\caption{ Model $M_1$ with one $1$-cycle}
\label{M1}
\end{subfigure}
\begin{subfigure}[t]{0.45\textwidth}
\centering
\begin{tikzpicture}[scale=2,every label/.style={color=blue}]
\coordinate [label=above:$x_1$] (x1) at (-0.4,0.688);
\coordinate [label=left:$x_2$] (x2) at (0,0);
\coordinate [label=above:$x_3$] (x3) at (0.4,0.688);
\coordinate [label=right:$x_4$] (x4) at (0.8,0);
\coordinate [label=right:$x_5$] (x5) at (1.2,0.688);
\draw [fill] (x1)circle[radius=.8pt];
\draw [fill] (x2)circle[radius=.8pt] -- (x3) circle[radius=.8pt]; 
\draw [fill] (x3)circle[radius=.8pt] -- (x4) circle[radius=.8pt];
\draw [fill] (x2) -- node[below] {$M_2$}(x4) circle[radius=.8pt];
\draw [fill] (x4)circle[radius=.8pt] -- (x5) circle[radius=.8pt];
\draw[pattern=north west lines, pattern color=blue](x2) -- (x3) -- (x4);
\end{tikzpicture}
\caption{$M_2$ is disconnected without cycles}
\label{M2}
\end{subfigure}
\caption{Two statistical models represented in terms of simplicial complex}
\label{M3}
\end{figure}

We consider two statistical models $M_1$ and $M_2$ variables $\mathbf{x}_1,\cdots,\mathbf{x}_5$ represented in terms of simplicial complex as in Figure \ref{M3}. In our setting,
vertices correspond to linear terms, edges to double interactions between variables and so on. Missing vertices or edges correspond to absent model terms. These multilinear hierarchical models can be seen as simplicial complexes.

The models can be written as 
\begin{align*}
E(Y_1|\tilde{X}_1)& = \mathbf{x}_1\theta_1 + \mathbf{x}_2\theta_2 +  \mathbf{x}_3\theta_3 +  \mathbf{x}_4\theta_4 + \mathbf{x}_5\theta_5 + \mathbf{x}_1\mathbf{x}_2\theta_{12} + \mathbf{x}_2\mathbf{x}_3\theta_{23} \\
& + \mathbf{x}_2\mathbf{x}_4\theta_{24} + \mathbf{x}_3\mathbf{x}_4\theta_{34} + \mathbf{x}_4\mathbf{x}_5\theta_{45}. \\
E(Y_2|\tilde{X}_2)& = \mathbf{x}_1\theta_1 + \mathbf{x}_2\theta_2 +  \mathbf{x}_3\theta_3 +  \mathbf{x}_4\theta_4 + \mathbf{x}_5\theta_5 +  \mathbf{x}_2\mathbf{x}_3\theta_{23}  + \mathbf{x}_2\mathbf{x}_4\theta_{24} \\
&+ \mathbf{x}_3\mathbf{x}_4\theta_{34} + \mathbf{x}_4\mathbf{x}_5\theta_{45} + \mathbf{x}_2\mathbf{x}_3\mathbf{x}_4\theta_{234}.
 \end{align*}
The model on the left in Figure \ref{M1} has a more complicated structure than the one on the right in Figure \ref{M2}:

the left model misses the interaction term $x_2x_3x_4$ which creates the
cycle (hole) framed by three interactions $x_2x_3$, $x_2x_4$ and $x_3x_4$.

The model on the right has the same size, yet it has a much simpler structure, without cycles (holes). Moreover it is collapsible in a topological sense \cite{LY2002}. These model descriptions can be reached by the Betti numbers using Homology groups in Section 2. It is not difficult to see the non-zero Betti numbers of the left model are $(1,1)$ which indicates that the simplicial complex associated to the regression model terms is connected and has one $1$-cycle, while for the right model are only $(2,0)$ which means the corresponding simplical complex is disconnected and there is no holes in the model structure. 

Betti numbers are used to describe the number of cycles for a simplicial complex as in previous section. We interpret the Betti numbers as the complexity of models which means if two different models have the same size of terms, the one has more Betti numbers is more complicated than the one with less Betti numbers as in the comparsion between $M_1$ and $M_2$. 

Since the selected nonzeros are not necessary to be hierarchical in the LASSO path, we pretend to force the non-zero coefficients into hierarchical structure at each break point of LASSO path by filling the missing terms. We use the corresponding simplicial complex formed by the original model terms together with missing terms to represent the topological structure of the regression models. Here, lower letter $x_i$ has been used to denote the vertex of the simplicial complex which corresponds to the variable $\mathbf{x}_i$ in the statistical regression models and the edge between $x_i$ and $x_j$ represents the interaction term $\mathbf{x}_i\mathbf{x}_j$.

\begin{exmp}\label{force_hie}
\normalfont{
For example, if the model has non-zero coefficients terms $ \mathbf{x}_1, \mathbf{x}_3, \mathbf{x}_1\mathbf{x}_2, \mathbf{x}_5\mathbf{x}_6$, we add extra terms in order to turn the model into a hierarchical one which becomes
$$ \mathbf{x}_1, \mathbf{x}_2, \mathbf{x}_1\mathbf{x}_2, \mathbf{x}_3, \mathbf{x}_5, \mathbf{x}_6, \mathbf{x}_5\mathbf{x}_6.$$
We can obtain the corresponding simplicial complex from the modified version of the coefficients and compute the Betti numbers which describe the number of cycles associated to the corresponding simplicial complex. In this example, there is only one Betti number $\beta_0 = 3$ which indicates there are three components in the simplicial complex formed by the non-zero terms. And there is no other cycles because the simplicial complex consists of a vertex and two edges. However, the model structure would become more complicated along with the increasing number of variables and the degree order of interactions.  
}
\end{exmp}

\section{Relationship between Betti numbers and Statistical features}
In the previous section, we show the simplicial complex can be applied to represent polynomial regression models. Now we would like to illustrate how the Betti numbers change over the LASSO path which in turn indicates the change process of the features in this section. This important finding provide the evidence that models with algebraic and topological representations can be potentially favoured by model selection process which are based on statistical criterion such as goodness of fit. 

Suppose there are $p$ variables which also can be interpreted as $p$ vertices of a simplicial complex. Pairwise interactions can be seen as edges in the simplicial complex, triple interactions are the $2$-simplex faces and so on. We focus on the case of $1$-cycles formed by edges and $2$-cycles by triangles, although the ideas we develop generalize naturally to higher-dimensional cycles.

\begin{figure}
\centering%
\begin{tikzpicture}[scale=5,every label/.style={color=blue}]
    \coordinate (a) at (4.5,0.86);
    \coordinate (b) at (4.5,0.3);
    \coordinate (c) at (5.3,0);
    \coordinate (d) at (4,0);
    \draw[very thick] (a) -- (c);
    \draw[very thick] (a) -- (d);
    \draw[very thick] (c) -- (d);
    \draw[thick, dash dot dot] (a) -- (b);
    \draw[thick, dash dot dot] (b) -- (d);
    \draw[thick, dash dot dot] (b) -- (c);
    \fill[black!20, draw=black, thick] (a) circle (1pt) node[black, above right] {$x_1$};
    \fill[black!20, draw=black, thick] (b) circle (1pt) node[black, above left] {$x_2$};
    \fill[black!20, draw=black, thick] (c) circle (1pt) node[black, below right] {$x_3$};
    \fill[black!20, draw=black, thick] (d) circle (1pt) node[black, above left] {$x_4$};
\end{tikzpicture}
\caption{A simplicial complex with four vertices, six edges.}
\label{all_edges}
\end{figure}
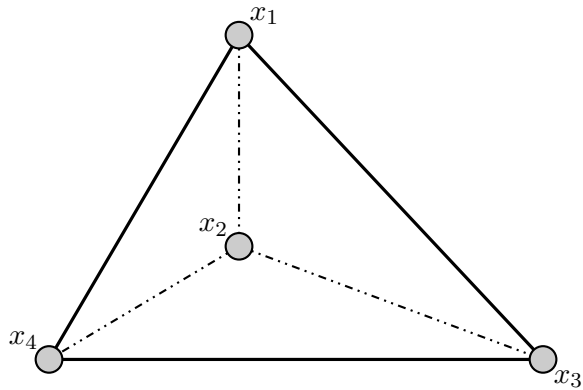

We start from the simplicial complex with four vertices and all edges as depicted in Figure \ref{all_edges}. By Lemma \ref{lemma4}, there are $\binom{4-1}{2}=3$ independent 1-cycles. Now we fill one 1-cycles, the number of independent $1$-cycles can be computed $\mbox{rank}(H_1)=\mbox{rank}(Z_1)-\mbox{rank}(B_1)= 1$ by the methods in Section \ref{2.2}.
This means we have only one independent 1-cycle after we fill two 1-cycles. By the same way, there would be no 1-cycles if we fill three 1-cycles and finally a 2-cycle (hollow tetrahedron) after filling all 1-cycles. The changes of the number of independent 1-cycles clearly indicate that higher-dimensional cycles are formed by paying the price of destroying of lower-dimensional cycles. We extend this idea to statistical models with high order interactions. The higher dimensional cycle is forming by destroying more independent lower dimensional cycles. Once the number of higher dimensional cycles achieve its maximal, the lower dimensional cycles would disappear.

\begin{figure}
  \includegraphics[width=\linewidth]{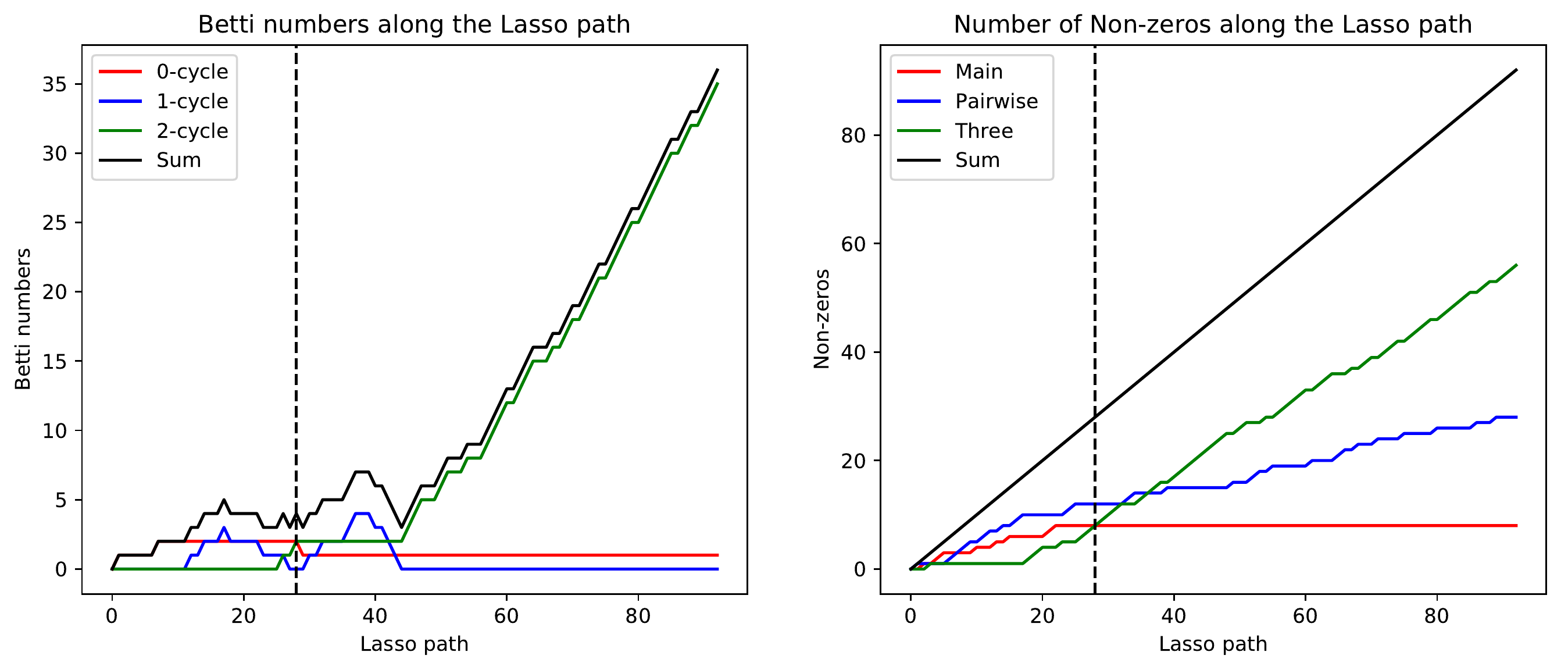}
  \caption{Plots illustrate the involving process of the Betti numbers and the number of non-zeros in a LASSO path. The Plot on the left are the Betti numbers of $0$-cycle, $1$-cycle, $2$-cycle and the sum of all Betti numbers over the LASSO path where the vertical line represent the true support (non-zero features). The plot on the right are the corresponding number of non-zero features of the main effects, pairwise interactions and three-order interactions. The vertical lines on both plots represent the true Betti numbers and non-zeros.}
  \label{fig:Betti changes}
\end{figure}

We can obtain the Betti numbers from each break point of the LASSO path as in Figure \ref{fig:Betti changes}. If we have large number of $\beta_{1}$, this means there are many non-zero pairwise interactions but may not lots of triple interaction features. However, a big value of $\beta_{2}$ means non-zero set of pairwise interactions and triple interactions as shown. To provide some intuition, The left panel of Figure \ref{fig:Betti changes} plots the change process of Betti numbers of $0$-cycle, $1$-cycle, $2$-cycle during the LASSO path for a true model which has $8$ variables, $12$ pairwise interactions and $8$ three-order interactions and the Betti number of the true model is $(\beta_0, \beta_1, \beta_2)=(2,0,2)$. The curve corresponds to $0$-cycle is relatively stable which indicates the number of components and  whether the model is connected. There is no $1$-cycles in the true model therefore the curve of $1$-cycle does not change too much though it is more turbulent than $0$-cycle curve. From one point later, all $1$-cycles are destroyed over the increasing of $2$-cycles. 
The right panel of Figure \ref{fig:Betti changes} depicts the number of non-zero coefficients corresponds to the main effects, pairwise interactions and three-order interactions. To some extent, the curves are consistent with the Betti number curves. As we could see that the curve of main effects is as stable as the curve of $0$-cycle, the number of pairwise interactions and three-order interactions are increasing because there are more $2$-cycles being generated during the process. From now on, the Betti numbers could be used to describe the change of model terms in regression models. The Betti numbers show the topological structure of the associated simplicial complex. However, we should point out that different statistical models may have the same Betti numbers, in other words, Betti numbers decide the uniqueness of the topological structures but the corresponding statistical models are not necessary the same. For example, the Betti numbers of the triangle formed by $x_1, x_2, x_3$ is $(\beta_0, \beta_1) = (1,1)$ and the corresponding statistical model is $\mathbf{x}_1, \mathbf{x}_2, \mathbf{x}_3, \mathbf{x}_1\mathbf{x}_2, \mathbf{x}_1\mathbf{x}_3, \mathbf{x}_2\mathbf{x}_3$. Obviously, another triangle with vertices $x_1, x_2, x_4$ has the same Betti numbers but represents a different model.

\section{Betti Numbers for Model selection}
We discussed the change process of Betti numbers along the LASSO path and show that the corresponding Betti numbers are changing consistently with the regression models. In this section, we would like to study whether the Betti numbers which count the number of independent cycles in the corresponding simplicial complex can be applied in model selection methods combining statistical criterion, namely mean-squared-errors.

To better present our method, we introduce the following notations for LASSO: The LASSO solution for hierarchical models:
\begin{align*}
\hat{\theta}^l_{\alpha}({\lambda}) = \arg \min_{{\theta}_{\alpha}} \frac{1}{2n}||Y-X\theta_{\alpha}||_2^2+\lambda||\theta_{\alpha}||_1 
\end{align*}
The set of predictor variables selected by the LASSO estimator $\hat{\theta}^l_{\alpha}(\lambda)$ is denoted by $\mathcal{M}_{\lambda}$,
\begin{align*}
\mathcal{M}_{\lambda} = \{\alpha \mid  \hat{\theta}_{\alpha}^{l}(\lambda)\neq 0 \}.
\end{align*}
where $\hat{\theta}_{\alpha}^{l}(\lambda)\neq 0$ is understood as component wisely.
We start with the standard LASSO. Its advantages are continuous and easy of computation. The main disadvantage is that LASSO estimators causes bias \cite{FL2001} and it tend to select more noise variables \cite{M2007} which cause more noise Betti numbers. In previous section, we illustrate that Betti numbers can not be used to deicide the statistical uniquely. A compromise between the statistical errors and the Betti numbers may help in the variable selection.
Therefore we consider to combine the weighted residual sum of squares and the sum of Betti numbers which we refer to as the compound criterion (CC) in our later sections.

We assume that the true model is of sparsity, i.e., the correct model is given by a part of the candidate model (say the first $q$ predictor variables). A noise variable is hence a variable with index larger than $q$. If there is any noise variable entering the LASSO estimator, the Betti number often is not the same as the true ones. This is because the selected nonzeros containing more noise variables means more complicated simplicial complex representations which leads to more noise Betti numbers. The idea behind the compound criterion is to penalize the noise Betti numbers caused by the noise variables so that we could improve the LASSO recovery performance. On the other side, if the number of nonzeros is much less than the true one in the LASSO regression, the mean-squared errors (MSE) of the LASSO estimator could be far bigger than the errors of the estimates near the true model. Hence the statistical criterion (MSE) would be the main criterion for the variable selection. As we know there may be higher order (triple or more) interactions instead of only double interactions. Our work would focus on higher order cases to explore the benefits of model selection by adding the algebraic criterion.

Let $p$ be the number of variables and $k$ be the order of $k$-interactions. We consider a regression model for an outcome variable $Y$ and predictors $\mathbf{x}_1,\cdots, \mathbf{x}_p$, with pairwise interactions, triple interactions until $k$-interactions of these predictors. In our context, We standardize data matrix $X$ so that its columns have mean $0$ and standard deviation $1$; we then form interaction terms from these standardized predictors and finally center the resulting columns to obtain new model matrix $X$ (we use the same notation here). This is the preparation for the standard hierarchical LASSO \cite{BST2015}. 

We split our data into train, validation and test set, $X_{tr}, X_{v}, X_{te}$. The training dataset is a dataset of examples used for learning, that is to fit the parameters of the regression model. The validation dataset is a dataset that is independent of the training dataset, but that follows the same probability distribution as the training dataset. The validation set is therefore a set of examples used only to assess the performance of the models learning from the training data which could help us choose the optimal. Finally, we apply the selected models to the prediction of the test dataset and choose the model with minimal model errors as we explained in (\ref{mode_error}).

The main idea of our method is the following:
First, we apply LASSO to the augmented training data matrix which consists of main effects and interaction terms with the corresponding repsonse variable $Y$. The coefficients at the break points can be obtained from the LASSO path. The LASSO coefficient is moving linearly between two break points over the shinkage factor $\lambda, \lambda \in [0, \infty)$ \cite{ET2004}. Moreover from Erfon et al \cite{ET2004}, the sign of the LASSO coefficients between two break points remains the same which indicates that we do not need to consider the change of coefficients over the LASSO path between any two break points. For a simple example, we assume that $P_1$ and $P_2$ are the end points of a piece in a LASSO another $P_3$ is the point between $P_1$ and $P_2$. Table \ref{tab_sign} summarizes all the possible cases of the sign of $P_3$ which is always same as either $P_1$ or $P_2$.

\begin{table}[h]
\caption{\label{tab_sign}Possible signs for $P_1, P_2$ and $P_3$}
\begin{center}
\begin{tabular}{ccc}
$P_1$&$P_3$&$P_2$\\\hline
+&+&+\\
+&+&0\\
0&+&+\\
0&-&-\\
-&-&0\\
-&-&-\\
0&0&0\\
\end{tabular}
\end{center}
\end{table}

Second, \texttt{python} is run by the command \texttt{SimplicialComplex} which has been built as a small function to compute the Betti numbers as follows:
 We extract the LASSO coefficient $\hat{\theta}^l_{\alpha}(\lambda)$ at each break point from the LASSO path and fill the missing terms as in section one such that it becomes hierarchical. We build the link between statistical models and topolocial simplicial complex so that the corresponding simplicial complex can be obtained from the nonzero features $\hat{\theta}^l_{\alpha}(\lambda)$. Finally we compute the Betti numbers by the command \texttt{SimplicialComplex} again in \texttt{python}. We then have a collection of LASSO estimates $\hat{\theta}^l_{\alpha}(\lambda)$ and  Betti numbers $b(\lambda)$, where $\lambda\in [0,\infty)$ is the shrinkage factor.  However, the LASSO shrinkage causes the estimates of the non-zero coefficients to be biased towards zero, and in general they are not consistent \cite{HT2001}. One way to reduce the bias is to identify the set of non-zero coefficients by LASSO, and then fit a linear model to the selected non-zero features which is known as LARS-OLS hybrid of Efron et al. (2004) \cite{ET2004}. The LARS-OLS estimator of $\mathcal{M}_{\lambda}$ is denoted by $\hat{\theta}^{lo}_{\alpha}(\lambda)$, namely "using Lars to find the model but not to estimate" \cite{ET2004}. In this case, all coefficients in the model $\mathcal{M}_{\lambda}$ are estimated by the OLS-solution.
\begin{align*}
\hat{\theta}^{lo}_{\alpha}(\lambda) = \arg \min_{\theta} \frac{1}{2n}||Y-X_{\mathcal{M}_{\lambda}}\theta_{\alpha}||_2^2.
\end{align*}
Therefore we reestimate the model coefficients by simple linear regression for the non-zero coefficients selected by LASSO at each break point. This can also be done by relaxed LASSO proposed in \cite{M2007} which pointed out relaxed LASSO and LARS-OLS have similar performance for sparse data.
The compound criteria we proposed consists of weighted test data errors and Betti numbers. We use it to select among candidate models by extracting the coefficients minimize the compound criterion. 
\begin{align} \label{equation1}
CC(\lambda, \mu)=(1-\mu)\frac{\Vert Y_v-X_v\hat{\theta}^{lo}_{\alpha}(\lambda)\Vert^2_2}{\underset{\lambda}{\max}\Vert Y_v-X_v\hat{\theta}^{lo}_{\alpha}(\lambda)\Vert^2_2}+\mu\frac{b(\lambda)A}{\underset{\lambda}{\max}\ b(\lambda)A}
\end{align}
where $A$ is a fixed vector to achieve the required sum of Betti numbers $\mu$ is the fixed weight between $0$ and $1$. For example, $A= \mathbf{1}$ where each element of $A$ is $1$ so that $b(\lambda)A$ gives us the sum of all Betti numbers. If the first element of $A$, $a_1 = 0$, then $b(\lambda)A$ would be the sum of Betti numbers without the $0$-cycle ones. Intuitively, the weight $\mu$ in (\ref{equation1}) measures the weight that is given to the Betti numbers. The percentage of the Betti numbers is to penalize the noise Betti numbers caused by noise variables which do not exist in the true models. The beauty of the principle is that by properly combining two 'extreme' cases one can obtain a 'compromise' estimator that performs better than either extreme in terms of the accuracy of prediction errors and model sparsity. We shall point out that the compound criterion varies over different dimensional cycles. In other words, lower and higher dimensional cycles can have significant differences for the model selection for example, the performance in terms of prediction errors can be very different for compound criteria using higher dimensional cycles and lower dimensional ones. Moreover, we can see that the it is the LARS-OLS that corresponds to the main effects of the three-stage of model selection process therefore the proportion of the statistical errors in the compound criteria should be overwhelming the Betti numbers part. In fact, our simulation results indicates that the ideal weight of Betti numbers is less than one percent, i.e. $\mu < 0.1$ which did not surprise us. Adding a small percentage of algebraic criterion could help improve the LASSO performance.

Based on the description above, we summarize our three-stage model selection method by the combination of statistical and algebraic criterion as follows.

\bigskip
\begin{algo}\label{algo_cc}{\normalfont Betti numbers in model selection by the compound criterion (\ref{equation1}) based on LASSO (\ref{equation2}).
\begin{itemize}
\item[\textbf{Input}] Data matrix $X_{tr}, X_{v}, X_{te}$ containing columns according to candidate model $M$ (no intercept), the corresponding response vector $Y_{tr}, Y_{v}, Y_{te}$, a grid of weight vector $\omega = \{\mu \mid 0= \mu_0 < \cdots <\mu_m = 1\}$, and the fixed vector $A$ (say $A$ = $\mathbf{1}$).
\item[\textbf{Output}] Selected model support $\mathcal{M}^*$ and estimates of coefficients in $\theta^*$.
\end{itemize}
\begin{itemize}
\item[\textbf{Step 1}] Compute all standard LASSO solutions with the Lars-algorithm in \cite{ET2004} under the LASSO modification using $X_{tr}$ and $Y_{tr}$. Let $\mathcal{M}_{\Lambda} = \{\mathcal{M}_{\lambda}\mid\lambda \in \Lambda \}$ be the resulting set of models and $\hat{\theta}^l_{\alpha}(\lambda) = \{\hat{\theta}^l_{\alpha}(\lambda) \mid \lambda \in \Lambda\}$ be the set of LASSO coefficients, where $\Lambda = \{0= \lambda_1 < \cdots <\lambda_N \}$ are the corresponding penalty values.
\item[\textbf{Step 2}] For every $\lambda$ in $\Lambda$, force $\mathcal{M}_{\lambda}$ into hierarchical structure $\overline{\mathcal{M}}_{\lambda}$ and compute the set of Betti numbers vector set $B_{\Lambda} = \{b(\lambda) | \lambda \in \Lambda \}$, where $b(\lambda)$ is a single Betti number vector of simplicial complex associated to $\overline{\mathcal{M}}_{\lambda}$. And then reestimate the set of LASSO coefficients $\hat{\theta}^l_{\alpha}(\lambda)$ to obtain LARS-OLS coefficients $\hat{\theta}^{lo}_{\alpha}({\Lambda}) = \{\hat{\theta}^{lo}_{\alpha}({\lambda})| \lambda \in \Lambda \}$.
\item[\textbf{Step 3}] For each $\lambda \in \Lambda$ and $\mu \in \omega$, compute the compound criterion, $CC = \{CC(\lambda, \mu)=(1-\mu)\frac{\Vert Y_v-X_v\hat{\theta}^{lo}_{\alpha}({\lambda})\Vert^2_2}{\underset{\lambda}{\max}\Vert Y_v-X_v\hat{\theta}^{lo}_{\alpha}({\lambda})\Vert^2_2}+\mu\frac{b(\lambda)A}{\underset{\lambda}{\max}\ b(\lambda)A}: \lambda \in \Lambda, \mu \in \omega\}$.
Finally, we pick the minimal compound error and the corresponding penalty value, that is 
\begin{align*}
\lambda^* = \arg \min_{\lambda\in \Lambda} \{\min_{\mu \in \omega}CC(\lambda, \mu)\} 
\end{align*}
so that $\mathcal{M}^* = \mathcal{M}_{\lambda^*}$, $\hat{\theta}^* = \hat{\theta}^{lo}_{\alpha}({\lambda^*})$.
\end{itemize}
}\end{algo}

It is necessary to know the computation complexity since the algorithm may look complicated at the first place. The computational complexity of the standard LASSO  is $O(n_{tr}n_p\mbox{min}\{n_{tr},n_p\})$, as there are $O(\mbox{min}\{n_{tr},n_p\})$ steps, each of complexity $O(n_{tr}n_p)$, where $n_{tr}$ is the size of train data set and $n_p$ is the size of candidate model. The computational complexity of LARS-OLS is $n_{tr}n_p^3$, here we require that $n_p \leq n_{tr}$. The computational complexity of the Betti numbers along the LASSO path is $O\big(\binom{p}{k+1}\binom{p}{k}(\binom{p}{k+1} + \binom{p}{k})k n_p\big)$ for candidate model having $d$-order interactions, $k+2\leq p$ by Theorem \ref{prop1}.

We use a simple example to better understand the algorithm above. We will only illustrate Step 1 and Step 2 in Algorithm \ref{algo_cc} and show more simulations results in the later Chapter.
\begin{exmp}\label{ex_alg}{\normalfont
This example illustrates the generation of
hierarchical models in the LASSO 
path in the first two steps of Algorithm 1.
We consider synthetic data given in Table \ref{tab_ex}
with $n=10$ design points in variables $x_1,x_2,x_3$ and
response variable $y$. 
\begin{description}
\item{\textbf{Input}}
A full squarefree model $M$ in three variables with no intercept 
was used with this data. The terms of $M$ were $x_1,x_2,x_3,x_1x_2,x_1x_3,
x_2x_3,x_1x_2x_3$. The design model matrix 
$X_{tr}$ was created from model $M$ and the data in Table \ref{tab_ex} using 
standard practice of centering (zero mean) and standardising (unit variance)
the columns for main factors $x_1,x_2,x_3$ while only centering interaction 
terms. To create $Y_{tr}$, the values of the response variable $y$ were 
centered and standardised.
\item{\textbf{Step 1}} Using the lars algorithm, the LASSO path was
computed. The path has $N=8$ breakpoint values $\lambda_1,\ldots,\lambda_8$ which are given
in the first column of Table \ref{tabex2}.
 Each row of the said table contains $\hat{\theta}^l_{\alpha}(\lambda)$, the LASSO estimates of
coefficients for the model $\mathcal{M}_\lambda$, given in the table as a complex.
\item{\textbf{Step 2}} 
For each $\mathcal{M}_\lambda$, the hierarchical closure $\overline{\mathcal{M}}_\lambda$, 
and the
first two Betti numbers of this closure were computed. They are 
given in the same table. 
Note that for simplicity, the complexes 
 $\mathcal{M}_\lambda$ and $\overline{\mathcal{M}}_\lambda$ in Table \ref{tabex2} 
do not contain the empty set.  
\end{description}

Consider the first row in Table \ref{tabex2}. The non-zero coefficients of
$\hat{\theta}^l_{\alpha}(\lambda)$ determine a complex $\mathcal{M}_\lambda$ which is not a simplicial complex as it does not have the terms 2 and 23 ($x_2$ and interaction $x_2x_3$). The closure $\overline{\mathcal{M}}_\lambda$ includes these terms so the complex is now a simplicial complex, corresponding to hierarchical model. In the second row, the complex is already a simplicial complex so the closure does not
add extra terms. It is until the fifth row that the closure requires adding extra terms. 

The column with Betti numbers $B(\overline{\mathcal{M}}_\lambda)$ in Table \ref{tabex2} has
the number of components of each complex (first entry) and one-dimensional cycles 
(second entry). Note that in this example, the rest of Betti numbers (number of 
two and higher-dimensional cycles) are zero.
Concerning the first Betti number, as $\lambda$ increases, the first 
three complexes are connected by interactions and thus they have only one component. 
It is only until the fourth complex \{1,2,3,12\} that two components appear due to lack of 
interaction between $x_3$ and the rest of model terms. Further down in the table the
number of components decrease to one in line seven as the coefficients of model terms 
have all shrank to zero apart from that for $x_3$. Regarding the number of one dimensional
cycles (second Betti number), only one complex has a cycle which appears due to an absence of 
triple interaction term $x_1x_2x_3$. This is shown in the second row of the Table \ref{tabex2}.

In the final row of Table \ref{tabex2}, when $\lambda=1.75$, all the model terms have shrank 
to zero. At this stage the model corresponds to the void complex, with zero Betti numbers.
}\end{exmp}

\begin{table}[h]
\caption{\label{tab_ex}Data for Example \ref{ex_alg}.}
\begin{center}
\begin{tabular}{cccc}
$x_1$&$x_2$&$x_3$&$y$\\\hline
0&0&0&1\\
0&1&0&1\\
1&1&1&1\\
1&1&1&0\\
0&1&1&0\\
0&0&1&0\\
1&0&0&0\\
0&0&1&0\\
1&1&0&1\\
1&0&1&0\\
\end{tabular}
\end{center}
\end{table}

\begin{table}
\caption{\label{tabex2}Models for the LASSO path of Example \ref{ex_alg}.}
\begin{center}
\scalebox{0.8}{
\begin{tabular}{|c|ccccccc|c|c|c|}\hline
\multicolumn{8}{|c|}{LASSO path with coefficient estimates $\hat{\theta}^l_{\alpha}(\lambda)$}&\multicolumn{3}{|c|}{Model, hierarchical closure and Betti numbers}\\
$\lambda$&1&2&3&12&13&23&123&$\mathcal{M}_\lambda$&$\overline{\mathcal{M}}_\lambda$&$B(\overline{\mathcal{M}}_\lambda)$\\\hline
0&-1.02&0&-1&1.94&1.94&0&-0.97&\{1,3,12,13,123\}&
\{1,2,3,12,13,23,123\}&(1,0)\\
0.06&-0.72&0.12&-0.83&1.23&1.13&-0.29&0&\{1,2,3,12,13,23\}&\{1,2,3,12,13,23\}&(1,1)\\
0.15&-0.48&0.08&-0.74&0.99&0.63&0&0&\{1,2,3,12,13\}&\{1,2,3,12,13\}&(1,0)\\
0.27&-0.15&0.14&-0.52&0.66&0&0&0&\{1,2,3,12\}&\{1,2,3,12\}&(2,0)\\
0.4&0&0.18&-0.46&0.31&0&0&0&\{2,3,12\}&\{1,2,3,12\}&(2,0)\\
0.97&0&0.08&-0.26&0&0&0&0&\{2,3\}&\{2,3\}&(2,0)\\
1.22&0&0&-0.18&0&0&0&0&\{3\}&\{3\}&(1,0)\\
1.75&0&0&0&0&0&0&0&\{\}&\{\}&(0,0)\\\hline
\end{tabular}}
\end{center}
\end{table}

\section{Model errors and modified AIC}\label{5.3}
We describe the prediction and model errors in computing the compound criteria and we also introduce an other variable selection method called non-negative garrote and the modified version of Akaike's information criterion (AIC) for the use in our later simulations.
\subsection{Prediction and Model Errors}

The prediction error is defined as the average error in the prediction of $Y$ given $X$ for future cases not used in the construction of a prediction equation. There are two regression situations, $X$ random and $X$ deterministic. However, in real problems there is often no clear distinction between the two scenarios, see \cite{T1996, FL2001}. Here we only consider the case of $X$ random for ease of presentation. 

In $X$ random cases, the data $(X, Y)$ are assumed to be a random sample from their parent distribution. Let $\hat{\eta}(X)$ be a prediction estimator constructed using the present data, the prediction error is defined as 
\begin{align}\label{pre_error}
\mbox{PE}(\hat{\eta}) = E\{Y - \hat{\eta}(X)\}^2,
\end{align}
where the expectation is taken only respect to the new observation $(X, Y)$. The prediction error can be decomposed as 
\begin{align}
\mbox{PE}(\hat{\eta}) = E\{Y - E(Y|X)\}^2 + E\{E(Y|X) - \hat{\eta}(X)\}^2
\end{align}
The first term, called the variance, measures the extent to which the predictions for individual data sets vary around their average. The second term is due to lack of fit to an underlying model called model error and denoted by $\mbox{ME}(\hat{\eta})$. Simulation results are reported in terms of $\mbox{ME}$ rather than $\mbox{PE}$. For the linear models $\hat{\eta}(X) = X\hat{\theta}$ considered in this paper, the model error has the sample form
\begin{align}\label{mode_error}
\mbox{ME}(\hat{\theta}) = (\hat{\theta}-\theta)^TE(XX^T)(\hat{\theta}-\theta).
\end{align}
It is worth pointing out that the oracle estimate, $\arg \min\{\mbox{ME}(\hat{\theta})\}$, is only computable in simulations not real examples.

For the linear models $\hat{\eta}(X) = X\hat{\theta}_{\alpha}$ considered in this section, the compound criterion in (\ref{equation1}) for our simulations becomes
\begin{align} \label{equation3}
CC(\lambda, \mu)=(1-\mu)\frac{\Vert (\hat{\theta}^{lo}_{\alpha}({\lambda})-\theta_{\alpha})^TE(X_v X_v^T)(\hat{\theta}^{lo}_{\alpha}({\lambda})-\theta_{\alpha})\Vert^2_2}{\underset{\lambda}{\max}\Vert (\hat{\theta}^{lo}_{\alpha}({\lambda})-\theta_{\alpha})^TE(X_v X_v^T)(\hat{\theta}^{lo}_{\alpha}({\lambda})-\theta_{\alpha})\Vert^2_2}+\mu\frac{b(\lambda)A}{\underset{\lambda}{\max}\ b(\lambda)A}
\end{align}
Again we point out that the oracle estimate, $\arg \min\{\mbox{ME}(\hat{\theta}_{\alpha})\}$, is only computable in simulations not real examples.

\subsection{Non-negative Garrote and Akaike's information criterion}
The original non-negative garrote estimator that was introduced  in \cite{B1995} is a scaled version of the least square estimate. The shrinking factor $D({\lambda}) = (d_1^{\lambda},\ldots, d_p^{\lambda})^T$, $d_j^{\lambda}>0$ is given as the minimizer to
\begin{align}\label{nng}
\hat{D}^{ng}({\lambda}) = \arg \min_{D} \frac{1}{2n}||Y-ZD||_2^2+\lambda||D||_1 
\end{align}

where $Z = (Z_1,\ldots, Z_p)$, $Z_j = \mathbf{x}_j\hat{\theta}^{ols}_j$ and $\hat{\theta}^{ols}$ is the original least square estimator based on (\ref{polire}). An efficient algorithm has been proposed in \cite{Y2007} to build the exact non-negative garrote solution path which is very similar to Lars algorithm in \cite{ET2004}. An alternative view of non-negative garrote is that by regarding $Z$ as the new design matrix, (\ref{nng}) can be solved as non-negative LASSO in \cite{W2014}. 

The second method is Akaike's information criterion (AIC) proposed by Akaike (1973) \cite{A1973}. AIC has been proposed to correct for the bias of maximum likelihood by the addition of a penalty term to compensate for the over-fitting of more complex models. Let $k$ be the number of estimated parameters in the regression model (\ref{polire}). Let $\hat{L}$ be the maximum value of the likelihood function for the model. Then the AIC value of the model is the following
\begin{align*}
-\mbox{ln}(\hat{L}) + k
\end{align*}
Under normal distributions, the likelihood is in fact the sum of square errors. In order to compare AIC with the compound criteria, we propose our modified AIC (MAIC) as follows
\begin{align}\label{AIC}
\mbox{MAIC}(\lambda, \mu)=(1-\mu)\frac{\Vert Y_v-X_v\hat{\theta}^{lo}_{\alpha}({\lambda})\Vert^2_2}{\underset{\lambda}{\max}\Vert Y_v-X_v\hat{\theta}^{lo}_{\alpha}({\lambda})\Vert^2_2}+\mu\frac{||\mathcal{M}_{\lambda}||_1}{\underset{\lambda}{\max}\ ||\mathcal{M}_{\lambda}||_1}
\end{align}
Similarly, the corresponding MAIC for our simulations becomes
\begin{align}\label{MAIC}
\mbox{MAIC}(\lambda, \mu)=(1-\mu)\frac{\Vert (\hat{\theta}^{lo}_{\alpha}({\lambda})-\theta_{\alpha})^TE(X_vX_v^T)(\hat{\theta}^{lo}_{\alpha}({\lambda})-\theta_{\alpha})\Vert^2_2}{\underset{\lambda}{\max}\Vert (\hat{\theta}^{lo}_{\alpha}({\lambda})-\theta_{\alpha})^TE(X_vX_v^T)(\hat{\theta}^{lo}_{\alpha}({\lambda})-\theta_{\alpha})\Vert^2_2}+\mu\frac{||\mathcal{M}_{\lambda}||_1}{\underset{\lambda}{\max}\ ||\mathcal{M}_{\lambda}||_1}
\end{align}

MAIC in (\ref{AIC}) which is the weighted combination of model errors and number of non-zero factors is in the same spirit of the compound criterion in (\ref{equation1}).

\section{ Numerical Experiments}
In this section, we report a simulation study done to compare the compound criterion algorithms with the LARS-OLS and several other statistical model selection methods. For the simulations we consider various linear models as in (\ref{polire}).
We split the data into train, validation and test set after standardizing the main effects and centring the interactions. In all examples, we used the LARS algorithm to compute the LASSO solutions. For each break point, we used \texttt{python} to compute the Betti numbers associated to the corresponding simplicial complex as we did in Chapter 3. We divide the interval $[0, 1]$ into $100$ equal length as the  weights of the compound criterion. The compound criterion consists of relaxed test errors of LARS-OLS coefficients and Betti numbers; thus the difference between LARS-OLS and modified version must be contributed by the weighted sum of Betti numbers. In the simulations, we use model errors instead of prediction errors in our simulations.  \\

\textbf{Outline of experiments} 

\begin{itemize}
\item Experiment 1 focuses on a group of comparison: four types of compound criteria and the LARS-OLS. We expect to find the advantages over LARS-OLS by using compound criteria. On the other side, we would like to explore which criterion would be better to put in real practice. The simulation results indicate that the compound criterion is of better prediction errors and less model complexity even when the noise level is high while have similar performance of mean-square-error but sparser model under small noise level. Finally, we numerically compare the optimal compound criterion with the LASSO, LARS-OLS \cite{EH2004}, non-negative garrote \cite{B1995} and modified AIC criteria in (\ref{AIC}). 
\item The recovery property is that the model selection method can recover the true non-zero variables from which we could obtain the true Betti numbers. Experiment 2 focuses on testing the performances of compound criteria in terms of prediction errors and selected number of coefficients but under more noise terms. This can be informally understood as testing the recovery property of the compound criterion which means to recover the true non-zero features of the regression models. This problem known variously as sparsity recovery, support recovery. Wainwright (2009) \cite{W2009} showed that LASSO could recover the true sparsity pattern at high probability when certain proper conditions are satisfied. In this experiment, we add more noise terms in the candidate model so that both of the prediction accuracy between the compound criterion and the LASSO may decline. The simulation procedure is basicly the same as in Experiment 1. Under more noise cases, our simulations indicate that the compound criteria lost its advantages for certain models and traditional variable selection methods and MAIC can be very useful for selecting models with less variables and prediction errors. 
\end{itemize}

\subsection{Experiment 1: Model with eight variables}

Four models were considered in the simulations and we focused on triple interactions among the predictors. In the first model we considered a hierarchical model including all main effects, pairwise interactions and one third of triple interactions. The second model is also hierarchical but disconnected with two components. For the third model, we fit an disconnected model with three components. The last model is a non-hierarchical model involving extra triple interactions without certain number of pairwise interactions. 

In all cases, design points in variables $\mathbf{x}_1,\cdots,\mathbf{x}_8$ realizations were simulated according to a centred multivariate normal distribution with covariance between $\mathbf{x}_i$ and $\mathbf{x}_j$ being $\rho^{|i-j|}$ for some value of $0\leq \rho <1$. We simulated 1000 datasets from the foregoing model. The regression noise $\epsilon$ is normally distributed with zero mean and standard deviation $\sigma=1, 3, 6$. And four compound criteria are built and compared. We refer to $0$-cycles (number of components) and $1$-cycles as lower dimensional cycles and the rest as higher dimensional cycles since our models focus on higher order interactions.  Four types of Betti number criteria are the total sum of Betti numbers $\sum_{i=0}\beta_i$, sum of Betti numbers without $0$-cycles $\sum_{i=1}\beta_i$, sum of Betti numbers without neither $0$-cycles or $1$-cycles $\sum_{i=2}\beta_i$, the sum of only higher dimensional cycles and sum of Betti numbers consisting of $0$-cycles or $1$-cycles $\beta_0 + \beta_1$. 

For each simulation, $625$ observations were collected and 60\% of these observations are used for training models, 20\% for validation set and the rest 20\% for the final comparison of the candidate models selected by the test data set. Coefficients for the regression model $\ref{polire}$ are the realizations from discrete uniform distribution $U(1,5)$ and the coefficients for each model are fixed in all the simulations. As standard in LASSO, the intercept is not considered. The candidate model contains full three degree interactions which consists of $\binom{8}{1}=8$ main factors, $\binom{8}{2}=28$ pairwise interactions and $\binom{8}{3}=56$ triple interactions formed by variables $\mathbf{x}_i\ (i=1,\cdots,8)$. We  centred the interactions so keep consistent with our algorithm in Section 4.
The response variable $Y$ was then simulated from \ref{polire}. On each dataset, we computed the entire solution path of the LASSO, then selected the coefficients of LASSO which minimize the compound criterion.  We compare the prediction accuracy of the LASSO and the compound criterion algorithm. The mean-squared error for test data of the estimator $X\hat{\theta}$ is given by the model error as in (\ref{equation3}) and (\ref{MAIC}). 
\begin{enumerate}[label=(\alph*)]
\item \textbf{Model 1}\  In model 1, there were all $8$ linear terms $X_1,\cdots,X_8$, all $\binom{8}{2}=28$ pairwise interactions and $10$ triple interactions therefore total $47$ terms in the true regression model.
 
\item  \textbf{Model 2}\  There were 31 variables in Model 2, this model was a hierarchical model and there were 8 main effects, 12 pairwise interactions and 8 triple interactions in Model 2. We refer to Model 2 as a disconnected model since the corresponding simplicial complex is disconnected with two components .
\item  \textbf{Model 3}\ Model 3 was also a hierarchical and disconnected model with three components including 8 main effects, 6 pairwise interactions and 6 triple interactions and total 20 terms .
\item  \textbf{Model 4}\ Model 4 is non-hierarchical containing 8 main effects, 12 pairwise interactions and 16 triple interactions, namely 36 variables in the true model.
\end{enumerate}
 Table \ref{table3} and Table \ref{table-cor} summarizes the average model error, model sizes in terms of the number of features selected under $\sigma_1=1$, $\sigma_2=3$ and $\sigma_3=6$ respectively when $\rho =0, 0.3$. The numbers in parentheses are standard deviations based on 1000 runs. The column labelled "All cycles" represents the criterion using sum of all Betti numbers, "No $0$-cycles" represents applying the sum of Betti numbers without $0$-cycles, "Higher cycles" means the sum of Betti numbers without neither $0$-cycles or $1$-cycles and "Lower cycles" represents the compound criterion of the sum of Betti numbers consisting of $0$-cycles and $1$-cycles. 
 
Several observations can be made from  Table \ref{table3} and Table \ref{table-cor}. The first three compound criteria ($\sum_{i=0}\beta_i$, $\sum_{i=1}\beta_i$,  and $\sum_{i=2}\beta_i$) including Betti numbers of higher cycles performs better than the one only includes lower dimensional cycles ($\beta_0 + \beta_1$) in all four models. The first two compound criteria ($\sum_{i=0}\beta_i$ and $\sum_{i=1}\beta_i$) often have similar performance in terms of prediction errors and selected number of nonzeros. In Model 1, the true Betti number vector corresponds to the associated simplicial complex is $(1,10,0,0,0,0)$ which indicates there is no higher dimensional cycles ($2$-cycles). By penalizing the potential $2$-cycles caused by the noise three-order interactions, the third type of compound criteria $\sum_{i=2}\beta_2$ performs best with better prediction errors and less non-zero factors among all of the methods. In Model 2 and Model 3, the true Betti number vector are $(2,0,2,0,0,0)$ and $(3,0,1,0,0,0)$ which do not contain any $1$-cycles, therefore the second compound criterion is applied to penalize the noise $1$-cycles and $2$-cycles. The corresponding Betti number vector of Model 4 is $(1,3,1,0,0,0)$ and  both the second compound criterion $\sum_{i=1}\beta_i$ and the third one $\sum_{i=2}\beta_i$ are suitable in this case.  The first criterion has the similar performance to the second one since the only difference between these two is $\beta_0$ which keeps almost constant in the LASSO path. The fourth type of criterion $\beta_1 + \beta_2$ has the worst performance in terms of prediction accuracy and the number of non-zero factors. This is because lower dimensional cycles are not enough to describe the models properly. We also notice that in most of the cases the minimal compound criterion happens when $\mu=0.05$. This means  only a small weight of the Betti numbers would lead to the improvement of prediction errors and reduction of the model complexity. 
\begin{sidewaystable}
\centering
\caption{Results for the four models that were considered in the simulations}
\label{table3}
\scalebox{0.9}{
\begin{tabular}{ccccccccccccccccc}
\toprule
& \multicolumn{16}{c}{Results for the following methods of candidate models with $3$-order interactions when $\rho = 0$}\\
\cline{2-16}
& \multicolumn{4}{c}{$\sigma_1 = 1$} & \multicolumn{4}{c}{$\sigma_2 = 3$} & \multicolumn{4}{c}{$\sigma_3 = 6$} \\  

\cmidrule(lr){2-5} \cmidrule{6-9} \cmidrule(lr){10-13} \cmidrule(lr){14-17} 
& $\sum_{i=0}\beta_i$ & $\sum_{i=1}\beta_i$ & $\sum_{i=2}\beta_i$ &  $\beta_0+\beta_1$ & $\sum_{i=0}\beta_i$ & $\sum_{i=1}\beta_i$ & $\sum_{i=2}\beta_i$ &  $\beta_0+\beta_1$ & $\sum_{i=0}\beta_i$ & $\sum_{i=1}\beta_i$ & $\sum_{i=2}\beta_i$ &  $\beta_0+\beta_1$\\ 
\midrule
Model 1 &   &   &  &   &   &  &  &  \\ 
 Model error & 0.24 & 0.24 &0.19& 0.35 &2.04 &2.04 & 2.03 &2.26 & 9.71 & 9.71 & 9.77 & 9.97 \\
  & (0.08) &(0.08) &(0.06) & (0.12) &0.77) &(0.77) & (0.79) &(0.85) &(3.46) & (3.45) &(3.51) &(3.58) \\ 
   \\ 
 Number of factors & 54.87 & 54.91 &50.28& 65.49 &53.6 &53.6 & 52.44 &56.7 & 54.64 & 54.65 & 53.06 & 56.88  \\ 
  & (2.96) &(2.95) &(2.19) & (5.47) &3.47) &(3.48) & (3.22) &(5.73) &(4.53) & (4.53) &(4.59) &(6.53) \\ 
 Model 2  \\ 
  Model error & 0.11 & 0.1 &0.1& 0.13 &1.22 &1.22 & 1.23 &1.26 & 6.43 & 6.43 & 6.53 & 6.53 \\ 
   & (0.04) &(0.04) &(0.04) & (0.08) &0.55) &(0.55) & (0.56) &(0.6) &(2.6) & (2.61) &(2.63) &(2.69)  \\ 
   Number of factors & 29.73 & 29.51 &29.71& 32.24 &31.71 &31.68 & 31.84 &32.22 & 31.76 & 31.75 & 31.89 & 32.76  \\ 
 & (1.44) &(1.6) &(1.73) & (7.52) &3.28) &(3.31) & (3.22) &(4.42) &(4.57) & (4.59) &(4.37) &(5.79)  \\
   Model 3   \\ 
Model error & 0.08 & 0.07 &0.07& 0.09 &0.76 &0.76 & 0.77 &0.8 & 4.39 & 4.38 & 4.46 & 4.43 \\  
& (0.03) &(0.03) &(0.03) & (0.05) &(0.38) &(0.39) & (0.41) &(0.41) &(2.02) & (2.02) &(2.08) &(2.08)  \\
 Number of factors & 21.46 & 20.65 &20.75& 22.96 & 21.93 &21.76 & 21.97 &22.24 & 22.08 & 21.91 & 22.28 & 22.41 \\
 & (0.96) &(0.96) &(1.09) & (5.42) &2) &(2) & (2.23) &(2.16) &(3.51) & (3.54) &(3.63) &(4.27)  \\
 Model 4  \\
 Model error & 0.14 & 0.14 &0.13& 0.23 &1.59 &1.59 & 1.59 &1.69 & 8.42 & 8.41 & 8.47 & 8.68  \\
 & (0.06) &(0.06) &(0.05) & (0.1) &0.69) &(0.69) & (0.7) &(0.73) &(3.23) & (3.24) &(3.36) &(3.41)  \\
 Number of factors & 39.71 & 39.72 &38.56& 48.34 &41.86 &41.87 & 41.53 &43.82 & 42.57 & 42.56 & 41.77 & 45.58  \\
 & (2.59) &(2.58) &(1.94) & (6.67) &3.42) &(3.42) & (3.39) &(5.11) &(5.39) & (5.38) &(5.48) &(7.08)   \\
   \bottomrule
  \multicolumn{17}{l}{Reported are the average model error, average number of factors in the selected  model over 1000 runs for compound criteria using}\\ 
\multicolumn{17}{l}{  different sum of Betti numbers.}
\end{tabular}}
\end{sidewaystable}

\begin{sidewaystable}
\caption{Results for the four models that were considered in the simulations}
\label{table-cor}
\scalebox{0.9}{
\begin{tabular}{ccccccccccccccccc}
\toprule
& \multicolumn{16}{c}{Results for the following methods of candidate models with $3$-order interactions when $\rho = 0.3$}\\
\cline{2-16}
& \multicolumn{4}{c}{$\sigma_1 = 1$} & \multicolumn{4}{c}{$\sigma_2 = 3$} & \multicolumn{4}{c}{$\sigma_3 = 6$} \\  

\cmidrule(lr){2-5} \cmidrule{6-9} \cmidrule(lr){10-13} \cmidrule(lr){14-17} 
& $\sum_{i=0}\beta_i$ & $\sum_{i=1}\beta_i$ & $\sum_{i=2}\beta_i$ &  $\beta_0+\beta_1$ & $\sum_{i=0}\beta_i$ & $\sum_{i=1}\beta_i$ & $\sum_{i=2}\beta_i$ &  $\beta_0+\beta_1$ & $\sum_{i=0}\beta_i$ & $\sum_{i=1}\beta_i$ & $\sum_{i=2}\beta_i$ &  $\beta_0+\beta_1$\\ 
\midrule
Model 1 &   &   &  &   &   &  &  &  \\ 
 Model error & 0.28 & 0.28 &0.19& 0.37 &1.92 &1.93 & 1.77 &2.68 & 7.93 & 7.93 & 7.85 & 8.57 \\
  & (0.11) &(0.11) &(0.06) & (0.14) &0.7) &(0.7) & (0.68) &(1.09) &(3.16) & (3.16) &(3.3) &(3.46) \\ 
   \\ 
 Number of factors & 56.86 & 56.87 &49.84& 66.08 &52.89 &52.93 & 50.47 &60.67 & 52.36 & 52.36 & 50.75 & 54.72  \\ 
  & (2.76) &(2.75) &(2.14) & (5.46) &3.42) &(3.42) & (2.27) &(5.94) &(3.43) & (3.42) &(2.89) &(5.3) \\ 
 Model 2  \\ 
  Model error & 0.11 & 0.1 &0.1& 0.13 &0.97 &0.96 & 0.98 &1.02 & 4.37 & 4.37 & 4.45 & 4.44 \\ 
   & (0.04) &(0.04) &(0.04) & (0.08) &0.41) &(0.41) & (0.43) &(0.46) &(2) & (1.99) &(2.08) &(2.03)  \\ 
   Number of factors & 29.58 & 29.16 &29.34& 31.26 &29.5 &29.39 & 29.62 &29.93 & 29.78 & 29.75 & 29.95 & 29.95  \\ 
 & (1.34) &(1.58) &(1.6) & (6.45) &1.46) &(1.52) & (1.74) &(2.53) &(2.05) & (2.09) &(2.25) &(2.24)  \\
   Model 3   \\ 
Model error & 0.08 & 0.07 &0.07& 0.12 &0.67 &0.66 & 0.67 &0.74 & 3.28 & 3.28 & 3.34 & 3.34 \\  
& (0.04) &(0.03) &(0.03) & (0.1) &(0.3) &(0.3) & (0.3) &(0.33) &(1.52) & (1.52) &(1.56) &(1.52)  \\
 Number of factors & 22.11 & 21.02 &21.13& 26.57 & 21.71 &21.5 & 21.62 &22.26 & 21.88 & 21.76 & 22.12 & 21.9 \\
 & (1.26) &(1.29) &(1.45) & (10.34) &1.49) &(1.53) & (1.68) &(1.65) &(2.14) & (2.16) &(2.45) &(2.05)  \\
 Model 4  \\
 Model error & 0.16 & 0.16 &0.13& 0.24 &1.34 &1.34 & 1.33 &1.79 & 6.23 & 6.24 & 6.26 & 6.67  \\
 & (0.07) &(0.07) &(0.05) & (0.1) &0.56) &(0.56) & (0.56) &(0.8) &(2.55) & (2.55) &(2.54) &(2.83)  \\
 Number of factors & 40.6 & 40.64 &38.63& 48.49 &40.02 &40.02 & 39.55 &45.12 & 40.17 & 40.18 & 39.62 & 42.51  \\
 & (2.96) &(2.98) &(2.04) & (5.9) &2.57) &(2.57) & (2.42) &(5.02) &(3.34) & (3.34) &(3.29) &(4.85)   \\
   \bottomrule
  \multicolumn{17}{l}{Reported are the average model error, average number of factors in the selected  model over 1000 runs for compound criteria using}\\ 
\multicolumn{17}{l}{  different sum of Betti numbers.}
\end{tabular}}
\end{sidewaystable}

In the above four models, we also numerically compare the optimal proposed variable selection methods in Table \ref{table-cor} with the LARS-OLS, non-negative garrote, LASSO, 5-fold Cross-validated LASSO and MAIC. We present the results of model errors and number of non-zero factors for the four models in Table \ref{Model1}, \ref{Model2}, \ref{Model3} and \ref{Model4}. In all four examples, the models that were selected by LASSO are larger than those selected by other methods (different partial sum of Betti numbers). This is to be expected since LASSO selected individual derived variables and, once a derived variables has been included in the model, the corresponding factor is present in the model. Therefore LASSO often produces unnecessarily large models in variable selection process. Also Cross-Validated LASSO tend to select more noise variables. Non-negative garrote has good performance of smaller model errors and selects less variables under low level noise for the four models. For example, non-negative garrote has even better model errors than the optimal compound criterion for Model 1. The compound criterion and MAIC have consistent good performance regardless of the noise level and the later one is often better than the former one with smaller model errors and less model complexity. Finally, LARS-OLS performs quite similarly to the optimal compound criterion because of the large proportion of compound errors in (\ref{equation1}) contributed by the statistical testing error from LARS-OLS.

\begin{table}
\centering
\caption{Results for Model 1 considered in the simulations}
\label{Model1}
\scalebox{1}{
\begin{tabular}{lcccccc}
\toprule
Method & \multicolumn{6}{c}{Moldel 1: Results for values of $\sigma$ when $\rho = 0.3$}\\
\cline{2-7}
& \multicolumn{2}{c}{$\sigma_1 = 1$} & \multicolumn{2}{c}{$\sigma_2 = 3$} & \multicolumn{2}{c}{$\sigma_3 = 6$} \\  
\cmidrule(lr){2-3} \cmidrule{4-5} \cmidrule(lr){6-7} 
& ME & Non-zeros & ME & Non-zeros & ME & Non-zeros \\
\midrule
CC       & 0.19  & 49.84  & 1.77  & 50.47  & 7.85 & 50.75  \\
         & (0.06)& (2.14) & (0.68) & (2.27) & (3.3) &(2.89)   \\   
Lars-OLS & 0.19  & 50.26  & 1.81  & 51  & 8.06 & 51.56 \\
         & (0.07)& (2.61) & (0.73) & (2.87) & (3.44) &(3.6) \\
LASSO-CV & 0.3  & 68.11 & 2.68  & 68.11 & 10.69 & 67.85  \\
         & (0.13) & (6.46) & (1.13) & (6.46) & (4.49) & (6.48) \\
NNG      & 0.18  & 48.95  & 2.01  & 52.39 & 10.4 & 56.76 \\
         & (0.06) & (2.6) & (0.77)& (3.73)& (4.25)& (5.42) \\
LASSO    & 0.3  & 68.12  & 2.69 & 68.12 & 10.69 & 67.86 \\
         & (0.13) & (6.49) & (1.13) & (6.49) & (4.49) & (6.52) \\
MAIC     & 0.18 & 49.41 & 1.77 & 50.03 & 7.76 & 50.14 \\
         & (0.06) & (1.91) & (0.72) & (2.18) & (3.28) & (2.86) \\
 \bottomrule
   \multicolumn{7}{l}{Reported are the average model error, average number of factors in the }\\ 
\multicolumn{7}{l}{selected model over 1000 runs for the statistical methods using CC, } \\
\multicolumn{7}{l}{ Lars-OLS, LASSO-CV, NNG, LASSO, MAIC } \\
\end{tabular}}
\end{table}

\begin{table}
\centering
\caption{Results for Model 2 considered in the simulations}
\label{Model2}
\scalebox{1}{
\begin{tabular}{lcccccc}
\toprule
Method & \multicolumn{6}{c}{Moldel 2: Results for values of $\sigma$ when $\rho = 0.3$}\\
\cline{2-7}
& \multicolumn{2}{c}{$\sigma_1 = 1$} & \multicolumn{2}{c}{$\sigma_2 = 3$} & \multicolumn{2}{c}{$\sigma_3 = 6$} \\  
\cmidrule(lr){2-3} \cmidrule{4-5} \cmidrule(lr){6-7} 
& ME & Non-zeros & ME & Non-zeros & ME & Non-zeros \\
\midrule
CC       & 0.1  & 29.16  & 0.96  & 29.39  & 4.37 & 29.75  \\
         & (0.04)& (1.58) & (0.41) & (1.52) & (1.99) &(2.09)   \\   
Lars-OLS & 0.1  & 29.36  & 0.98  & 29.67 & 4.5 & 30.19  \\
         & (0.04)& (1.64) & (0.43) & (1.8) & (2.14) &(2.42) \\
LASSO-CV & 0.2  & 49.29 & 1.79  & 49.29 & 7.14 & 49.22  \\
         & (0.08) & (6.97) & (0.73) & (6.97) & (2.89) & (7.01) \\
NNG      & 0.11  & 29.69  & 1.23  & 32.97 & 6.79 & 36.83 \\
         & (0.04) & (2.55) & (0.56)& (3.79)& (3)& (5.32) \\
LASSO    & 0.2  & 49.29  & 1.79 & 49.29 & 7.14 & 49.23 \\
         & (0.08) & (6.97) & (0.73) & (6.97) & (2.89) & (7.01) \\
MAIC     & 0.1 & 28.91 & 0.95 & 29.23 & 4.33 & 29.55 \\
         & (0.04) & (1.21) & (0.41) & (1.43) & (1.98) & (1.9) \\
 \bottomrule
   \multicolumn{7}{l}{Reported are the average model error, average number of factors in the }\\ 
\multicolumn{7}{l}{selected model over 1000 runs for the statistical methods using CC, } \\
\multicolumn{7}{l}{ Lars-OLS, LASSO-CV, NNG, LASSO, MAIC } \\
\end{tabular}}
\end{table}

\begin{table}
\centering
\caption{Results for Model 3 considered in the simulations}
\label{Model3}
\scalebox{1}{
\begin{tabular}{lcccccc}
\toprule
Method & \multicolumn{6}{c}{Moldel 3: Results for values of $\sigma$ when $\rho = 0.3$}\\
\cline{2-7}
& \multicolumn{2}{c}{$\sigma_1 = 1$} & \multicolumn{2}{c}{$\sigma_2 = 3$} & \multicolumn{2}{c}{$\sigma_3 = 6$} \\  
\cmidrule(lr){2-3} \cmidrule{4-5} \cmidrule(lr){6-7} 
& ME & Non-zeros & ME & Non-zeros & ME & Non-zeros \\
\midrule
CC       & 0.07  & 21.02  & 0.66  & 21.5  & 3.28 & 21.76  \\
         & (0.03)& (1.29) & (0.3) & (1.53) & (1.52) &(2.16)   \\   
Lars-OLS & 0.07  & 21.13  & 0.67  & 21.62  & 3.35 & 22.17 \\
         & (0.03)& (1.46) & (0.3) & (1.68) & (1.57) &(2.48) \\
LASSO-CV & 0.15  & 40.59 & 1.33  & 40.59 & 5.32 & 40.52  \\
         & (0.06) & (6.61) & (0.55) & (6.61) & (2.2) & (6.6) \\
NNG      & 0.07  & 21.42  & 0.83  & 24.17 & 4.7 & 27.52 \\
         & (0.03) & (2) & (0.41)& (3.15)& (2.13)& (4.3) \\
LASSO    & 0.15  & 40.59  & 1.33 & 40.59 & 5.32 & 40.52 \\
         & (0.06) & (6.61) & (0.55) & (6.61) & (2.2) & (6.6) \\
MAIC     & 0.07 & 20.82 & 0.66 & 21.33 & 3.25 & 21.49 \\
         & (0.03) & (1.11) & (0.3) & (1.42) & (1.49) & (2.04) \\
 \bottomrule
   \multicolumn{7}{l}{Reported are the average model error, average number of factors in the }\\ 
\multicolumn{7}{l}{selected model over 1000 runs for the statistical methods using CC, } \\
\multicolumn{7}{l}{ Lars-OLS, LASSO-CV, NNG, LASSO, MAIC } \\
\end{tabular}}
\end{table}

\begin{table}
\centering
\caption{Results for Model 4 considered in the simulations}
\label{Model4}
\scalebox{1}{
\begin{tabular}{lcccccc}
\toprule
Method & \multicolumn{6}{c}{Moldel 4: Results for values of $\sigma$ when $\rho = 0.3$}\\
\cline{2-7}
& \multicolumn{2}{c}{$\sigma_1 = 1$} & \multicolumn{2}{c}{$\sigma_2 = 3$} & \multicolumn{2}{c}{$\sigma_3 = 6$} \\  
\cmidrule(lr){2-3} \cmidrule{4-5} \cmidrule(lr){6-7} 
& ME & Non-zeros & ME & Non-zeros & ME & Non-zeros \\
\midrule
CC       & 0.13  & 38.63  & 1.33  & 39.55  & 6.26 & 39.62  \\
         & (0.05)& (2.04) & (0.56) & (2.42) & (2.54) &(3.29)   \\   
Lars-OLS & 0.13  & 38.91  & 1.35  & 39.97  & 6.41 & 40.4 \\
         & (0.05)& (2.26) & (0.58) & (2.8) & (2.73) &(4.03) \\
LASSO-CV & 0.24  & 59.45 & 2.15  & 59.44 & 8.48 & 58.99  \\
         & (0.09) & (7.24) & (0.82) & (7.24) & (3.19) & (7.27) \\
NNG      & 0.13  & 38.09  & 1.56  & 42.19 & 8.4 & 45.77 \\
         & (0.05) & (2.56) & (0.65)& (4.01)& (3.22)& (6.13) \\
LASSO    & 0.24  & 59.45  & 2.15 & 59.45 & 8.48 & 58.99 \\
         & (0.13) & (6.49) & (1.13) & (6.49) & (4.49) & (6.52) \\
MAIC     & 0.18 & 49.41 & 1.77 & 50.03 & 7.76 & 50.14 \\
         & (0.09) & (7.24) & (0.82) & (7.25) & (3.19) & (7.28) \\
 \bottomrule
   \multicolumn{7}{l}{Reported are the average model error, average number of factors in the }\\ 
\multicolumn{7}{l}{selected model over 1000 runs for the statistical methods using CC, } \\
\multicolumn{7}{l}{ Lars-OLS, LASSO-CV, NNG, LASSO, MAIC } \\
\end{tabular}}
\end{table}

\subsection{Experiment 2: Model with eight variables but more noise terms}
We consider the same models and conditions as in Experiment 1 except that the candidate model now contains full four degree interactions which is comprised of  $\binom{8}{1}=3$ main factors, $\binom{8}{2}=28$ pairwise interactions, $\binom{8}{3}=56$ triple interactions and $\binom{8}{4}=70$ quadruple interactions formed by variables $\mathbf{x}_i\ (i=1,\cdots,8)$. 

Table \ref{table4} summarizes the average model errors, model complexity as we did in Experiment 1. Comparing to Table \ref{table-cor}, the candidate model with more noisy terms has worse prediction errors and the identified model is more complicated than the counterpart. The optimal compound criteria in each model agrees with the ones when the candidate models have three-order interactions. We also compare the optimal compound criterion with other methods as we did in Experiment 1. To our surprise, non-negative garrote performs best for having smallest model errors and selecting least non-zero factors under low noise level when $\sigma = 1$. MAIC has slightly bigger model errors and model complexity than non-negative garrote. The performance of optimal compound criterion is quite similar to LARS-OLS for often having the same model errors and similar selected model size. LASSO and Cross-validated LASSO tend to have worse model errors and select more noise terms as in Experiment 1. However, we could still observe the benefits of the compound criteria and MAIC under the higher level noise when $\sigma =3, 6$. Both the compound criteria and MAIC provide relatively consistent results while non-negative garrote is no longer the best one but having greater model errors and selecting more noise terms. This behavior which contrasts dramatically with different sparsity pattern in Table \ref{table3} and \ref{table4} raises an interesting question as to whether there exist a threshold that the compound criteria could produce consistent results.
\begin{sidewaystable}
\centering
\caption{Results for the four models that were considered in the simulations}
\label{table4}
\scalebox{0.9}{
\begin{tabular}{ccccccccccccccccc}
\toprule
& \multicolumn{16}{c}{Results for the following methods of candidate models with $4$-order interactions when $\rho = 0.3$}\\
\cline{2-16}
& \multicolumn{4}{c}{$\sigma_1 = 1$} & \multicolumn{4}{c}{$\sigma_2 = 3$} & \multicolumn{4}{c}{$\sigma_3 = 6$} \\  

\cmidrule(lr){2-5} \cmidrule{6-9} \cmidrule(lr){10-13} \cmidrule(lr){14-17} 
& $\sum_{i=0}\beta_i$ & $\sum_{i=1}\beta_i$ & $\sum_{i=2}\beta_i$ &  $\beta_0+\beta_1$ & $\sum_{i=0}\beta_i$ & $\sum_{i=1}\beta_i$ & $\sum_{i=2}\beta_i$ &  $\beta_0+\beta_1$ & $\sum_{i=0}\beta_i$ & $\sum_{i=1}\beta_i$ & $\sum_{i=2}\beta_i$ &  $\beta_0+\beta_1$\\ 
\midrule
Model 1 &   &   &  &   &   &  &  &  \\
 Model error & 0.27 & 0.27 &0.23& 0.33 &2.27 &2.27 & 2.23 &2.64 & 10.27 & 10.27 & 10.28 & 10.89 \\
  & (0.16) &(0.16) &(0.11) & (0.16) &0.98) &(0.98) & (0.96) &(1.24) &(4.71) & (4.7) &(4.87) &(5.16) \\
   \\
 Number of factors & 58.48 & 58.55 &56.02& 62.78 &57.33 &57.34 & 56.56 &60.47 & 57.62 & 57.62 & 56.85 & 59.73  \\
  & (8.23) &(8.44) &(5.62) & (5.76) &4.23) &(4.22) & (4.41) &(4.82) &(4.88) & (4.89) &(4.88) &(5.33) \\
 Model 2  \\
  Model error & 0.12 & 0.12 &0.12& 0.13 &1.14 &1.14 & 1.14 &1.22 & 5.47 & 5.47 & 5.5 & 5.64 \\
   & (0.06) &(0.06) &(0.06) & (0.08) &0.56) &(0.56) & (0.56) &(0.62) &(2.77) & (2.76) &(2.76) &(2.87)  \\
   Number of factors & 32.42 & 32.3 &32.24& 33.24 &32.58 &32.55 & 32.56 &33.34 & 33.24 & 33.22 & 33.18 & 33.85  \\
 & (2.81) &(2.94) &(2.7) & (4.21) &2.88) &(2.92) & (2.8) &(3.73) &(3.14) & (3.44) &(3.44) &(3.33)  \\
   Model 3   \\
Model error & 0.09 & 0.08 &0.08& 0.11 &0.79 &0.79 & 0.8 &0.9 & 3.94 & 3.94 & 3.99 & 4.11 \\
& (0.04) &(0.04) &(0.04) & (0.08) &(0.39) &(0.49) & (0.49) &(0.49) &(1.84) & (1.85) &(1.9) &(1.99)  \\
 Number of factors & 23.75 & 23.04 &23.14& 25.34 & 23.97 &23.85 & 23.89 &24.78 & 24.04 & 23.98 & 24.19 & 24.48 \\
 & (2.2) &(2.33) &(2.37) & (4.87) &2.63) &(2.66) & (2.65) &(3.28) &(3.16) & (3.18) &(3.21) &(3.63)  \\
 Model 4  \\
 Model error & 0.17 & 0.17 &0.16& 0.2 &1.64 &1.64 & 1.63 &1.85 & 7.91 & 7.9 & 7.88 & 8.51  \\
 & (0.09) &(0.09) &(0.08) & (0.11) &0.78) &(0.78) & (0.78) &(0.96) &(3.76) & (3.74) &(3.73) &(4.42)  \\
 Number of factors & 42.85 & 42.92 &42.22& 45.35 &43.48 &43.48 & 43.23 &45.64 & 43.06 & 43.06 & 42.81 & 45.32  \\
 & (4.42) &(4.74) &(4.41) & (4.05) &3.65) &(3.66) & (3.7) &(4.02) &(4.34) & (4.34) &(4.47) &(5.01)   \\
   \bottomrule
  \multicolumn{17}{l}{Reported are the average model error, average number of factors in the selected  model over 1000  runs for compound criteria using}\\ 
\multicolumn{17}{l}{  different sum of Betti numbers.}
\end{tabular}}
\end{sidewaystable}

\begin{table}
\centering
\caption{Results for Model 1 with full four-order interactions}
\label{Model1-4dm}
\scalebox{1}{
\begin{tabular}{lcccccc}
\toprule
Method & \multicolumn{6}{c}{Moldel 1: Results for values of $\sigma$ when $\rho = 0.3$}\\
\cline{2-7}
& \multicolumn{2}{c}{$\sigma_1 = 1$} & \multicolumn{2}{c}{$\sigma_2 = 3$} & \multicolumn{2}{c}{$\sigma_3 = 6$} \\
\cmidrule(lr){2-3} \cmidrule{4-5} \cmidrule(lr){6-7}
& ME & Non-zeros & ME & Non-zeros & ME & Non-zeros \\
\midrule
CC       & 0.23& 56.02 & 2.23  & 56.56  & 10.28 & 56.85  \\
         & (0.11)& (5.62) & (0.96)  & (4.41)  & (4.87) & (4.88)   \\
Lars-OLS & 0.23 & 56.07 & 2.31  & 57.39  & 10.83 & 58.2 \\
       & (0.09) & (4.25) & (1.04)  & (4.54)  & (5.32) & (5.5) \\
LASSO-CV & 0.43& 81.38 & 3.87  & 81.38  & 15.34 & 80.97  \\
      & (0.19) & (9.6) & (1.72)  & (9.6)  & (6.79) & (9.75) \\
NNG      & 0.19 & 50.29 & 2.68  & 57.34  & 17.84 & 64.68\\
         & (0.07) & (3.66) & (1.27)  & (5.65)  & (8.87) & (7.86) \\
LASSO    & 0.43& 81.38 & 3.87  & 81.38  & 15.34 & 80.97 \\
         & (0.19) & (9.6) & (1.72)  & (9.61)  & (6.79) & (9.75) \\
MAIC     & 0.22& 54.94 & 2.2  & 56.06  & 10.1 & 56.37 \\
         & (0.08) & (3.7) & (0.94)  & (4.08)  & (4.57) & (4.69) \\
 \bottomrule
   \multicolumn{7}{l}{Reported are the average model error, average number of factors in the }\\
\multicolumn{7}{l}{selected model over 1000 runs for the statistical methods using CC, } \\
\multicolumn{7}{l}{ Lars-OLS, LASSO-CV, NNG, LASSO, MAIC } \\
\end{tabular}}
\end{table}

\begin{table}
\centering
\caption{Results for Model 2 with full four-order interactions}
\label{Model2-4dm}
\scalebox{1}{
\begin{tabular}{lcccccc}
\toprule
Method & \multicolumn{6}{c}{Moldel 2: Results for values of $\sigma$ when $\rho = 0.3$}\\
\cline{2-7}
& \multicolumn{2}{c}{$\sigma_1 = 1$} & \multicolumn{2}{c}{$\sigma_2 = 3$} & \multicolumn{2}{c}{$\sigma_3 = 6$} \\
\cmidrule(lr){2-3} \cmidrule{4-5} \cmidrule(lr){6-7}
& ME & Non-zeros & ME & Non-zeros & ME & Non-zeros \\
\midrule
CC       & 0.12& 32.3 & 1.14  & 32.56  & 5.47 & 33.22  \\
         & (0.06)& (2.94) & (0.56)  & (2.8)  & (2.76) & (3.44)   \\
Lars-OLS & 0.12 & 32.02 & 1.16  & 32.8  & 5.65 & 33.71 \\
       & (0.05) & (2.75) & (0.58)  & (3.91)  & (2.89) & (3.73) \\
LASSO-CV & 0.26& 58.21 & 2.37  & 58.22  & 9.42 & 58.14  \\
      & (0.12) & (10.24) & (1.05)  & (10.24)  & (4.19) & (10.37) \\
NNG      & 0.11 & 30.78 & 1.7  & 37.37  & 11.09 & 43.03\\
         & (0.05) & (3.07) & (0.91)  & (5.72)  & (5.6) & (7.57) \\
LASSO    & 0.26& 58.22 & 2.37  & 58.21  & 9.42 & 58.13 \\
         & (0.12) & (10.24) & (1.05)  & (10.24)  & (4.19) & (10.37) \\
MAIC     & 0.11& 31.38 & 1.11  & 32.06  & 5.33 & 32.7 \\
         & (0.05) & (2.32) & (0.55)  & (2.63)  & (2.65) & (3.21) \\
 \bottomrule
   \multicolumn{7}{l}{Reported are the average model error, average number of factors in the }\\
\multicolumn{7}{l}{selected model over 1000 runs for the statistical methods using CC, } \\
\multicolumn{7}{l}{ Lars-OLS, LASSO-CV, NNG, LASSO, MAIC } \\
\end{tabular}}
\end{table}

\begin{table}
\centering
\caption{Results for Model 3 with full four-order interactions}
\label{Model3-4dm}
\scalebox{1}{
\begin{tabular}{lcccccc}
\toprule
Method & \multicolumn{6}{c}{Moldel 3: Results for values of $\sigma$ when $\rho = 0.3$}\\
\cline{2-7}
& \multicolumn{2}{c}{$\sigma_1 = 1$} & \multicolumn{2}{c}{$\sigma_2 = 3$} & \multicolumn{2}{c}{$\sigma_3 = 6$} \\
\cmidrule(lr){2-3} \cmidrule{4-5} \cmidrule(lr){6-7}
& ME & Non-zeros & ME & Non-zeros & ME & Non-zeros \\
\midrule
CC       & 0.08& 23.04 & 0.79  & 23.85  & 3.94 & 23.98  \\
         & (0.04)& (2.33) & (0.49)  & (2.66)  & (1.85) & (3.18)   \\
Lars-OLS & 0.08 & 23.01 & 0.8  & 23.96  & 4.07 & 24.52 \\
       & (0.03) & (2.39) & (0.49)  & (2.78)  & (1.98) & (3.5) \\
LASSO-CV & 0.19& 47.58 & 1.72  & 47.58  & 6.82 & 47.31  \\
      & (0.08) & (9.15) & (0.73)  & (9.15)  & (2.85) & (9.2) \\
NNG      & 0.07 & 22.39 & 1.14  & 27.93  & 7.49 & 33.03\\
         & (0.04) & (2.79) & (0.68)  & (4.74)  & (3.72) & (6.55) \\
LASSO    & 0.19& 47.58 & 1.72  & 47.58  & 6.82 & 47.32 \\
         & (0.08) & (9.15) & (0.73)  & (9.15)  & (2.85) & (9.2) \\
MAIC     & 0.07& 22.52 & 0.78  & 23.51  & 3.83 & 23.53 \\
         & (0.03) & (1.96) & (0.48)  & (2.39)  & (1.72) & (3.03) \\
 \bottomrule
   \multicolumn{7}{l}{Reported are the average model error, average number of factors in the }\\
\multicolumn{7}{l}{selected model over 1000 runs for the statistical methods using CC, } \\
\multicolumn{7}{l}{ Lars-OLS, LASSO-CV, NNG, LASSO, MAIC } \\
\end{tabular}}
\end{table}

\begin{table}
\centering
\caption{Results for Model 4 with full four-order interactions}
\label{Model4-4dm}
\scalebox{1}{
\begin{tabular}{lcccccc}
\toprule
Method & \multicolumn{6}{c}{Moldel 4: Results for values of $\sigma$ when $\rho = 0.3$}\\
\cline{2-7}
& \multicolumn{2}{c}{$\sigma_1 = 1$} & \multicolumn{2}{c}{$\sigma_2 = 3$} & \multicolumn{2}{c}{$\sigma_3 = 6$} \\
\cmidrule(lr){2-3} \cmidrule{4-5} \cmidrule(lr){6-7}
& ME & Non-zeros & ME & Non-zeros & ME & Non-zeros \\
\midrule
CC       & 0.16& 42.22 & 1.63  & 43.23  & 7.88 & 42.81  \\
         & (0.08)& (4.41) & (0.78)  & (3.7)  & (3.73) & (4.47)   \\
Lars-OLS & 0.15 & 42.24 & 1.69  & 43.99  & 8.37 & 44.48 \\
       & (0.07) & (3.49) & (0.87)  & (4.14)  & (4.29) & (5.27) \\
LASSO-CV & 0.34& 70.1 & 3.03  & 70.1  & 11.88 & 69.38  \\
      & (0.16) & (10.69) & (1.4)  & (10.7)  & (5.45) & (10.48) \\
NNG      & 0.14 & 39.4 & 2.32  & 47.81  & 14.31 & 52.07\\
         & (0.06) & (3.58) & (1.25)  & (6.92)  & (7.16) & (8.51) \\
LASSO    & 0.34& 70.11 & 3.03  & 70.09  & 11.88 & 69.39 \\
         & (0.16) & (10.7) & (1.4)  & (10.7)  & (5.45) & (10.47) \\
MAIC     & 0.15& 41.44 & 1.6  & 42.97  & 7.81 & 42.53 \\
         & (0.06) & (2.97) & (0.76)  & (3.5)  & (3.73) & (4.44) \\
 \bottomrule
   \multicolumn{7}{l}{Reported are the average model error, average number of factors in the }\\
\multicolumn{7}{l}{selected model over 1000 runs for the statistical methods using CC, } \\
\multicolumn{7}{l}{ Lars-OLS, LASSO-CV, NNG, LASSO, MAIC } \\
\end{tabular}}
\end{table}

\section{Real Data}
To illustrate our results further, we analyse the red wine data set from the study \cite{CP2009}. This data set consists of 1599 sample of red wine from Portugal which was used to have wine quality test. The response variable is the quality score between $0$ and $10$. The predictors are eleven attributes of the red wine:fixed acidity (facid), volatile acidity (vacid), citric acid (cacid), residual sugar (rsugar), chlorides (chlo), free sulfur dioxide (fsdio), total sulfur dioxide (tsdio), density, pH, sulphates (sulp) and alcohol. 

\begin{figure}
\centering
  \includegraphics[width=.55\linewidth]{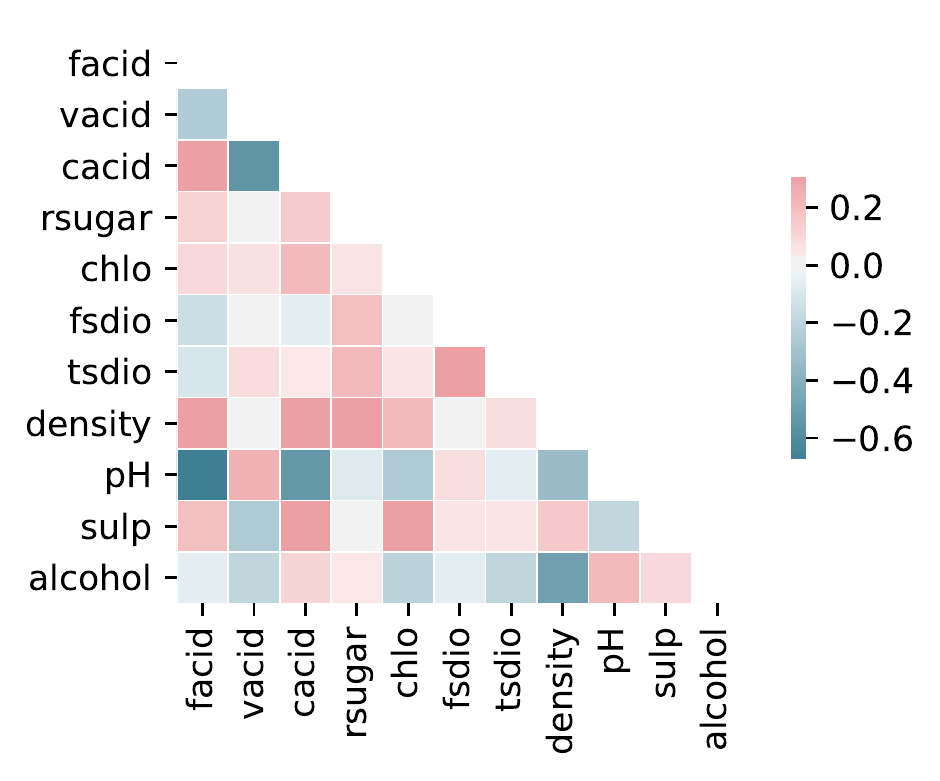}
  \caption{The correlation matrix of the training data set with $11$ variables.}
  \label{fig:corr-matrix}
\end{figure}

One the main interests here is to identify which predictors are more important in predicting the response and to find out if there exist higher order interactions among all the predictors. Figure \ref{fig:corr-matrix} gives the pairwise correlations between any two variables among the predictors. There exist significant pairwise correlations among the predictors, for example pH is significantly negative correlated to facid while positive correlated to alcohol. We thus reasonably further assume that there may exist three-order interactions instead of only pairwise interactions. For illustration, we randomly split the data set into three parts which are used for training (64\%), validating (16\%) and testing (20\%) respectively. The candidate model is the polynomial regression model containing full three-order interaction of the eleven main variables so that there are total $\binom{11}{1} + \binom{11}{2} + \binom{11}{3} = 231$ terms without the intercept. For the preparation of our methods, we first standardize the data set and then form the interactions as we did in Algorithm \ref{algo_cc}. We report the prediction errors and the selected factors in Table \ref{Tab:SRNRValues} by using the compound criteria, LASSO-OLS, Cross-validated LASSO, Non-negative garrote, LASSO and MAIC as we did in the simulations previously. As we repeat the random partition of the data set for ten times, the results are quite similar.
\begin{table}[ht]
\caption{Test set prediction error of the models selected by compound criteria, Lars-OLS, LASSO-CV, Non-negative garrotte, LASSO and MAIC}
\label{Tab:SRNRValues}
\begin{center}
\begin{tabular}{|l|c|c|c|}
\hline
Method & Prediction error & Non-zero factors \\
 \hline
CC($\sum_{i=0}\beta_i$) & 0.45 & 13     \\
CC($\sum_{i=1}\beta_i$) & 0.45 & 13  \\
CC($\sum_{i=2}\beta_i$) & 0.48 & 26    \\
CC($\beta_1 + \beta_2$) & 0.58 & 120     \\
Lars-OLS & 0.61 &  90 \\
LASSO-CV & 0.54 & 100 \\
NNG      &  0.47 & 80   \\
LASSO    &  0.54 & 100   \\
MAIC     &  0.44 & 16    \\
\hline
\end{tabular}
\end{center}
\end{table}

From Table \ref{Tab:SRNRValues}, the compound criterion $\sum_{i=1}\beta_i$ (same as $\sum_{i=0}\beta_i$) and MAIC perform the best which have better prediction accuracy and select less model terms as well. Even Non-negative garrote has performance of perdiction accuracy, it tends to select more noise vairables. Both of LASSO-CV and standard LASSO have the same performance in terms of prediction errors and the size of non-zero factors which is worse than Non-negative garrote. The compound criterion $\beta_0+\beta_1$ has the worst performance for selecting the most redundant variables. The selected important variables by the compound criterion are vacid, sulp, alcohol, pH and tsdio  which agrees with findings in \cite{CP2009}.

\section{Conclusions}
In this paper, we have proposed the compound criteria which combine statistical errors and topological Betti numbers for model selection of polynomial regression models containing high-order interaction. The main motivation are the shortcomings of LASSO which often selects more noise variables, especially applying cross-validation and additionally has low accuracy of prediction errors caused by the presence of many noise variables. In our setting, there are four types of compound criterion due to the different partial sum of Betti numbers which are used to penalize noise cycles caused by noise variables in the LASSO process. We introduce a concrete way to compute the Betti numbers by the operations of the corresponding matrices. A key advantage of our method is that it makes use of the topological structure uniqueness (to some extent) of the true model together with statistical criterion (MSE) to have better model selection results. To be specific, we compare our method to LARS-OLS to show the performance of the compound criteria. The advantages of our method are twofold: First, the number of selected coefficients is sparser than LARS-OLS without compromising on the accuracy of predictions. The models selected by the compound criterion is of less complexity. Second, it has more accurate prediction errors. If the noise level is very low, the predictive accuracy of both LASSO-OLS and compound criteria is comparable. However, our method has better performance than LARS-OLS when the noise level is very high.

\begin{appendices}

\section{Proofs}
\subsection{Proof of Lemma \ref{lemma2}}

For any $(d+1)$-simplex $\sigma$ in the simplicial complex $\Delta$, we show that $\partial_d\partial_{d+1}\sigma=0$. The boundary of $\sigma$, $\partial_{d+1}\sigma$ consists of all $d$-simplex of $\sigma$. So the boundary of $\partial_{d+1}\sigma$, $\partial_d\partial_{d+1}\sigma=0$ consists of all $(d-1)$-faces of $\partial_{d+1}\sigma$ belonging to exactly two $d$-faces. Hence, $\partial_d\partial_{d+1}\sigma=0$.

\subsection{Proof of Theorem \ref{prop1}}
A $d$-cycle must have at least $d+2$ vertices since a single $d$-simplex contains $d+1$ vertices. In order to show that $\Omega^d_{d+2}$ is a $d$-cycle, we just need to show that any $(d-1)$-simplex of $\Omega^d_{d+2}$ belongs to exact two its $d$-simplex. By the Definition \ref{d-complete}, we have 
\begin{align*}
\Omega^d_{d+2} = \sum_{r=1}^{d+2}[x_1,\ldots,x_{r-1},x_{r+1},\ldots,x_{d+2}]
\end{align*}
Taking the boundary, this gives us
\begin{align*}
\partial \  \Omega^d_{d+2} = \sum_{r=1}^{d+2}\sum_{\substack{i=1, \\ i\neq r}}^{d+2}[x_1,\ldots, x_{i-1}, x_{i+1}, \ldots,x_{r-1},x_{r+1},\ldots,x_{d+2}]
\end{align*}
We notice that $[x_1,\ldots, x_{i-1}, x_{i+1}, \ldots,x_{r-1},x_{r+1},\ldots,x_{d+2}]$ is contained in both 

$[x_1,\ldots,x_{r-1},x_{r+1},\ldots,x_{d+2}]$ and $[x_1,\ldots,x_{i-1},x_{i+1},\ldots,x_{d+2}]$ but not any other $d$-simplexes, therefore we have $\partial \  \Omega^d_{d+2} = 0$

On the other side, $\sigma$ is any $d$-cycle on $d+2$-vertices, there are only $d+2$ possible $d$-simplex on a set of $d+2$ vertices. We want to show that these $d+2$ $d$-simplex are distinct. Let $F$ be any $d$-simplex of $\sigma$ and $f$ be one of its $(d-1)$-simplex. We know that $f$ must belong to at least one other $d$-simplex since $\sigma$ is a $d$-cycle. We notice that there is only one vertex $v$ of $\sigma$ not contained in $F$. So the simplex formed by $f \cup \{v\}$ must be 
$d$-simplex of $\sigma$. Since there are $d+1$ different $(d-1)$-simplex which must long to another $d$-simplex of $\sigma$, this gives rise to $d+1$ distinct $d$-simplex which all contain $v$. Therefore $\sigma$ contains $d+2$ $d$-simplex including $F$, this proves $\sigma = \Omega^d_{d+2}$.

\subsection{Proof of Lemma \ref{lemma4}}

We know the simplest $1$-cycle is a triangle formed by three vertices. In order to count the independent cycles, we fix the first vertex of this triangle so there are $\binom{d+1}{2}$ for the rest two places. It is obvious that these $\binom{d+1}{2}$ cycles are independent. Next we prove that other 1-cycles can be represented by the subset of these independent cycles. 

Without losing generality, we fix the fist vertex of the independent cycles above by $x_1$. Any triangle containing $x_1$ belongs to the independent cycles set. For a 1-cycle with vertices $x_ix_jx_k, 1<i<j<k\leq d+2$, the cycle can be written as $\sigma=[x_ix_j]+[x_ix_k]+[x_jx_k]$. Therefore $\sigma$ can be represented by other independent 1-cycles, namely
\begin{align*}
& \sigma=\sigma_1+\sigma_2+\sigma_3 \\
& \sigma_1=[x_1x_i]+[x_1x_j]+[x_ix_j] \\
& \sigma_2=[x_1x_i]+[x_1x_k]+[x_ix_k] \\
& \sigma_3=[x_1x_j]+[x_1x_k]+[x_jx_k]
\end{align*}

\subsection{Proof of Theorem \ref{pa1}}
We use the same method as in Lemma \ref{lemma4}. The simplest $k$-cycle is the one formed by $k+2$ vertices as illustrated in Theorem \ref{prop1} . Without losing generality, we fix the fist vertex $x_1$ of this triangle so there are $\binom{d+1}{k+1}$ for the rest $k+1$ places. Again these $\binom{d+1}{k+1}$ cycles are independent. In the same way, we can prove that any $k$-cycle in the simplicial complex can be represented by a subset of these independent $k$-cycles.

\end{appendices}


\end{document}